# A Label-Free Physics-to-Data Acceleration Framework for Parametric Time-Dependent PDEs with Latent-Space Differential-Operator Learning

Hongjiang WANG, Weizhe WANG*, Yingzheng LIU

**Corresponding author (Email: wangwz0214@sjtu.edu.cn; Tel/Fax: +86-21-34205083)*

*Shanghai Jiao Tong University, Shanghai 200240, China*

## Abstract

Efficiently solving time-dependent parametric partial differential equations (PDEs) is a central task in computational science and engineering. Deep-learning-based accelerators span a physics-to-data spectrum, ranging from physics-informed solvers with strong physical consistency but high computational cost to data-driven surrogates with efficient inference but strong dependence on high-fidelity datasets. Although methods across this spectrum have been extensively studied, they have largely evolved in isolation and are often only loosely connected through explicit solution-field labels. In this work, we propose PHD-SF, a compact physics-to-data spectrum framework for accelerating time-dependent parametric PDEs. Through the SD–TPD separation strategy and the differential-operator learning strategy, PHD-SF enables reusable model information, including spatial features, latent dynamical features, and spatial differential-operator information, to be generated, inherited, and enriched across three operating modes—PIDON, HIDON, and DIDON—within a unified DON-based architecture. This design realizes compact label-free spectrum modeling without relying on precomputed full-field solution labels or explicit solution-label transfer among the operating modes. Furthermore, PHD-SF establishes a separated-solving–hyper-reduction-like-enrichment–direct-inference acceleration path, avoiding the dependence on large-scale high-fidelity data generation and its high offline computational cost in conventional label-mediated workflows. Results on four benchmark time-dependent PDE problems demonstrate accurate cross-parameter solving and direct inference using only one or two representative parameter cases in the initial physics-informed solving stage. The total end-to-end computational cost is lower than that required to train a conventional PINN for a single parameter case, while enabling reusable cross-parameter inference. PHD-SF therefore provides a new efficient solution path for many-query parametric PDE.

## Introduction

Dynamical systems are ubiquitous in science and engineering, and their evolution is commonly governed by time-dependent partial differential equations (PDEs) defined over space and time. A central task is to repeatedly solve such PDEs over broad parameter ranges and under diverse initial and boundary conditions, in support of online monitoring, system control, design optimization, and digital twins[1-3]. Although high-fidelity numerical solvers are reliable, each parameter case typically requires independent spatial discretization and time marching, making many-query and real-time computation prohibitively expensive. Developing fast and scalable acceleration methods for parametric PDEs is therefore a longstanding challenge in computational science.

Acceleration of parametric PDEs, however, is not a new problem. Reduced-order modelling (ROM) has long been a mainstream paradigm, supported by fairly mature theory and algorithms. Early studies on high-dimensional flow identification and energetic mode extraction developed Proper Orthogonal Decomposition (POD) [4, 5]. Building on POD, Galerkin-projection ROMs became a standard implementation of projection-based ROMs (PROMs) [6-8]. The Reduced Basis Method (RBM) further introduced an offline–online decomposition and certified error estimation, enabling multi-query and real-time solutions for parametric PDEs [8]. For large-scale nonlinear systems, the Empirical Interpolation Method (EIM) and its discrete variant DEIM reconstruct nonlinear terms from a small set of representative degrees of freedom [9, 10]. The idea of operator reconstruction from sparsely sampled degrees of freedom has driven the development of hyper-reduced-order models (HROMs) [1, 11-13], which typically achieve orders-of-magnitude speedups over full-order numerical solvers while preserving physical consistency and controllable errors. In this sense, the numerical-ROM community has already developed a relatively coherent theory of acceleration for parametric PDEs.

By contrast, deep-learning-based acceleration models (DL-AMs) for parametric PDEs have developed rapidly in recent years, resulting in a broad family of data-driven, hybrid, and physics-constrained approaches. Among data-driven models, non-intrusive ROMs such as POD–NN construct surrogate mappings from input parameters to latent states, enabling fast online inference[14], while nonlinear-manifold ROMs exploit encoder–decoder architectures to learn low-dimensional nonlinear solution manifolds[13, 15]. Beyond these approaches, operator-learning methods learn mappings between function spaces and can be viewed as resolution-invariant black-box surrogates for parametric PDEs, exhibiting strong expressive power and acceleration potential on complex parametric problems [16]. Representative examples include operator learners grounded in universal approximation theorems, such as DeepONet [17, 18], and integral-transform-based neural operators, such as the Fourier Neural Operator (FNO) [19]. However, purely data-driven models often suffer from high data requirements and limited physical consistency. Recent studies have shown that injecting physical knowledge into these fast solvers can effectively reduce data demand and enhance physical interpretability, thereby striking a better balance among data cost, physical consistency and computational speed [20]. Representative PINN-based approaches include physics-reinforced neural networks[21] and GPT-PINNs [22], whereas representative operator-learning models include PI-DeepONet [18], Physics-Informed Neural Operators [23] and physics-informed deep-learning ROMs (PI-DL-ROMs) [24, 25]. On the high-fidelity acceleration side, separable network architectures combined with forward-mode automatic differentiation (AD) have achieved computational complexity that scales linearly with the problem dimension, leading to orders-of-magnitude speedups for deep-learning-based PDE solvers [26-28]. Overall, these DL-AMs have demonstrated strong potential in non-intrusive implementation, nonlinear feature extraction, direct inference, inverse modeling, knowledge distillation, and transfer learning, while also driving the study of parametric PDE acceleration beyond traditional numerical ROMs toward a broader and more diverse family of deep learning frameworks [2, 29, 30].

Taken together, both classical numerical ROMs and emerging DL-AMs have produced a variety of methods with distinct characteristics. Viewed through the lens of physics-constraint strength and data dependence, these approaches can be organized along a continuous spectrum. At one end lie strongly physics-constrained and weakly data-dependent models with high computational cost, such as finite element method and PINN. At the other end lie weakly physics-constrained but strongly data-dependent models with extremely low online inference cost, such as DeepONet-type surrogates. Between these extremes are hybrid models, including HROMs and physics-reinforced neural networks[21]. Although methods across the physics-to-data spectrum have attracted substantial attention, they have largely evolved within separate research communities. The connections among them are often discussed only at the level of loose conceptual analogies or explicit data transfer. A typical workflow is that an upstream model, such as a high-fidelity solver, first generates labeled solution data, and a downstream data-driven model then learns a low-dimensional latent representation or a latent-state surrogate mapping from these labels. From a more structural perspective, however, such latent representations or surrogate mappings may already be identified during the upstream solving process itself, rather than being relearned afterwards through a separate supervised training stage based on explicitly generated solution labels.

Therefore, the relationships among models across the physics-to-data spectrum should not be understood merely as the transfer of explicit solution labels. A deeper connection lies in the reusable model information formed during the solving process and subsequently inherited or enriched by downstream models. This information includes representation-related quantities, such as spatial features, latent dynamical features and differential information, as well as structure-related information, such as the dependence of conventional HROMs on the discrete structures of their corresponding high-fidelity models [31]. If such information can be generated, inherited, and further enriched within a unified architecture, the transition from strongly physics-constrained solving, to hybrid modeling, and finally to fast direct inference can be realized without relying entirely on external labeled datasets or independent supervised training procedures. This perspective shifts DL-AMs from isolated acceleration models toward systematic PDE-acceleration frameworks based on cross-stage information reuse.

To realize such cross-stage information reuse, a unified computational and architectural framework is required. We observe that the mature and coherent acceleration theory developed in numerical ROMs, together with the flexible computational architectures of deep learning, provides a natural basis for unified modeling. Along this line, we propose PHD-SF, a compact physics-to-data spectrum framework for accelerating parametric time-dependent PDEs. PHD-SF consists of three operating modes within a unified differential operator network based (DON-based) architecture: physics-informed differential operator network (PIDON), physics-and-data hybrid informed DON (HIDON), and data-driven DON (DIDON). More fundamentally, these should be understood as three operating modes of a single DON-based architecture across different modes of the physics-to-data spectrum (a more detailed interpretation is provided in Appendix S1). Unlike conventional workflows in which upstream solvers generate solution-field labels for independently trained downstream surrogates, PHD-SF treats the solving process itself as a source of reusable model information, where solution and feature learning are performed simultaneously. Its central innovation lies in a cross-stage mechanism for the generation, inheritance, and enrichment of reusable model information within a unified DON-based architecture. Specifically, during physics-informed solving, PIDON generates reusable spatial features and the corresponding spatial differential information through separable differential-operator learning. HIDON then inherits these quantities and enriches latent dynamical features over a broader parameter domain at low cost by enforcing governing-equation residuals on sparse spatial collocation points. DIDON further reuses the inherited spatial features and latent dynamical features to enable fast direct inference. In this way, the transition from strongly physics-constrained solving, to hybrid enrichment, and finally to data-informed inference is realized not through external solution-label transfer, but through the internal generation, inheritance, and enrichment of reusable model information across the physics-to-data spectrum. Under this design, PHD-SF not only reduces the computational burden of repeated case-by-case optimization in conventional PINN-type methods, but also alleviates the dependence of purely data-driven surrogates on large-scale high-fidelity solution labels, thereby providing a compact solution path for many-query parametric time-dependent PDEs that balances physical consistency, computational efficiency, and cross-parameter inference capability.

We evaluate PHD-SF on four benchmark time-dependent PDE problems using a sparse, one-shot-like parameter configuration. Specifically, PIDON performs physics-constrained solving on only one or two representative parameter cases, HIDON enriches the latent dynamics over a broader parameter domain, and DIDON conducts large-scale direct inference. The results show that PHD-SF achieves accurate physics-constrained solutions for representative cases and reliable cross-parameter inference, while its total end-to-end computational cost is substantially lower than that required to train a conventional PINN for a single parameter case.

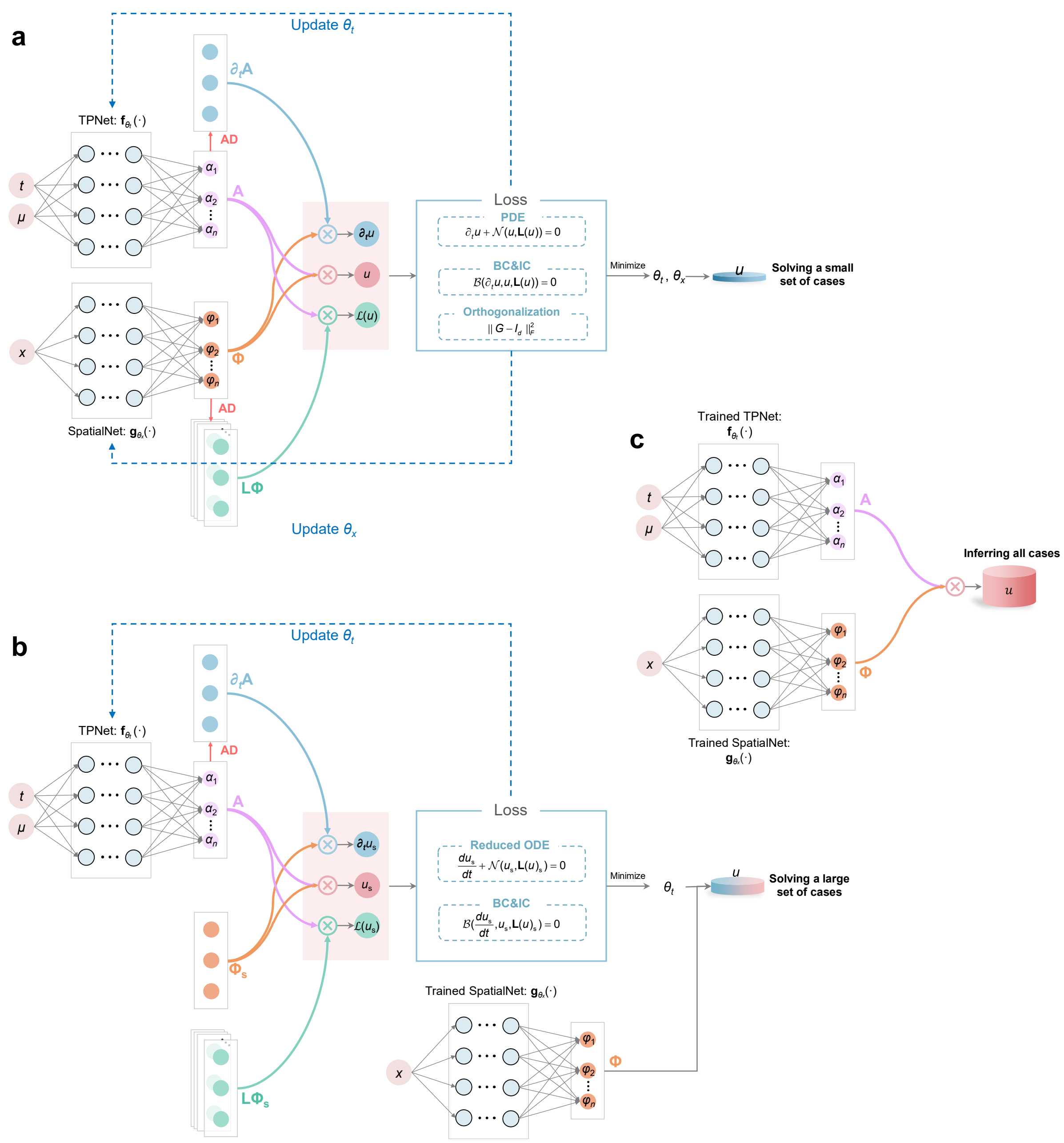

**Fig. 1 | Architectures of (a) PIDON, (b) HIDON and (c) DIDON.** PIDON couples a TPNet that outputs BCFs $\mathbf{A}(t,\boldsymbol{\mu})$ with a SpatialNet that outputs BFs $\boldsymbol{\Phi}(\boldsymbol{x})$. Temporal and spatial derivatives ($\partial_t\mathbf{A}$, $\mathbf{L}\boldsymbol{\Phi}$) are obtained by AD and combined via an inner product to form the PDE residual loss and update both subnetworks. HIDON inherits the learned BFs, SDOs, and SpatialNet, evaluates the residual on a sparse set of control points, and updates

only TPNet to enrich the BCFs. DIDON directly inherits the trained SpatialNet and TPNet and performs inference.

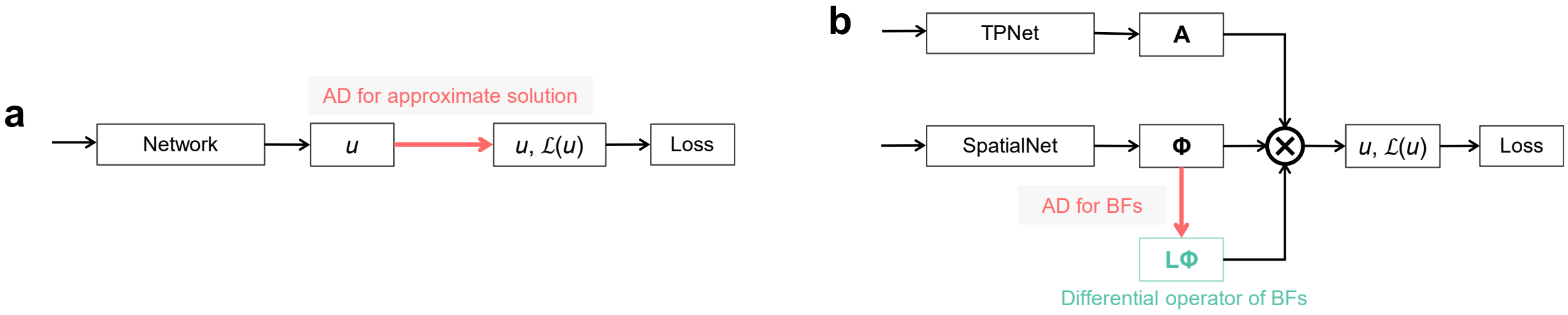

**Fig. 2 | Schematic comparison of differentiation mechanisms in conventional learning-based PDE solvers and the proposed PIDON. a**, Conventional solvers apply AD directly to the approximate solution $u$ to evaluate loss function. **b,** PIDON applies AD to the learned BFs to obtain the associated differential operators, and evaluates loss function indirectly.

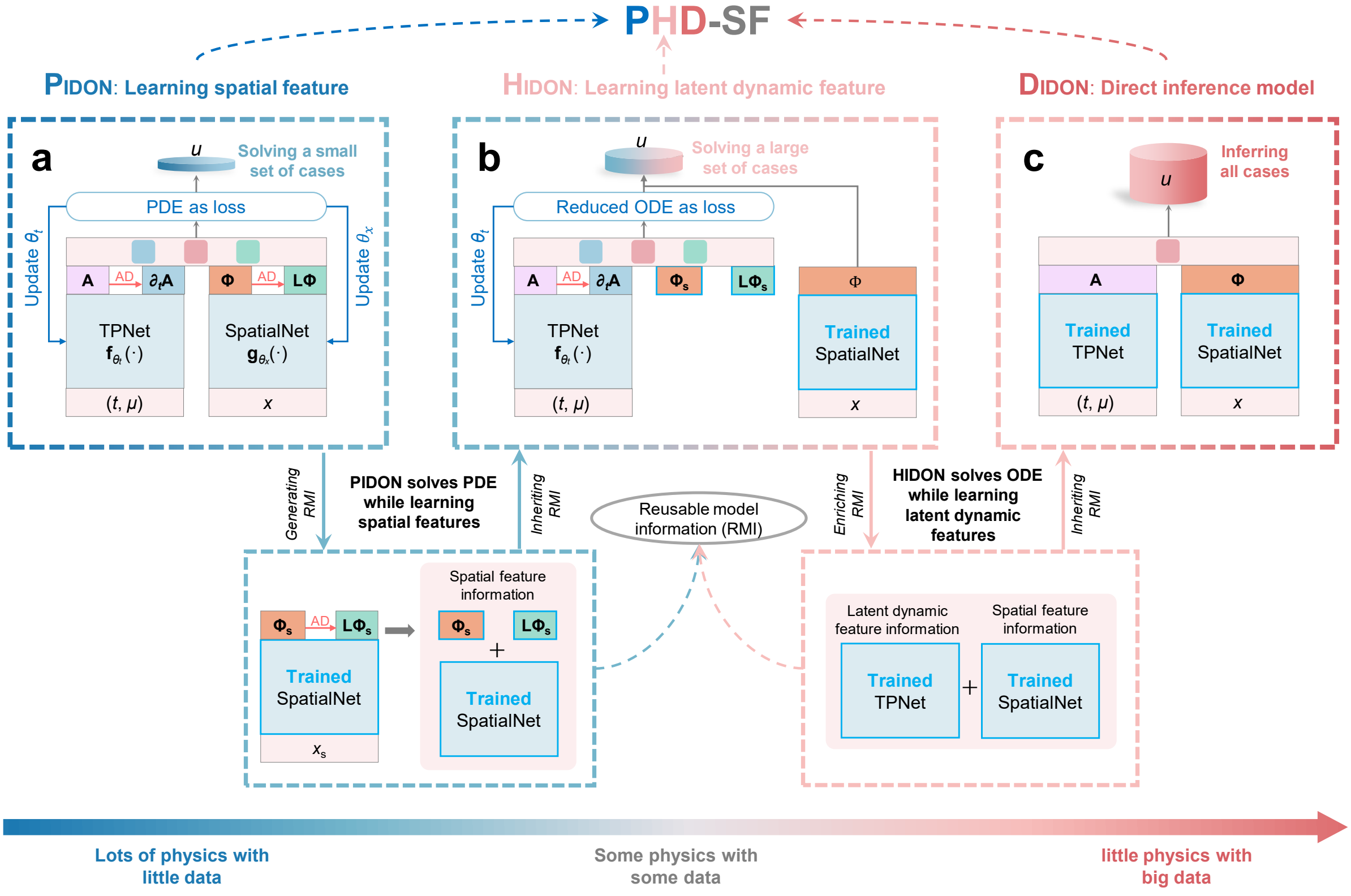

**Fig. 3| Overview of PHD-SF across the physics-to-data spectrum models.** PHD-SF enables reusable model information to be generated, inherited, and enriched across three operating modes—PIDON, HIDON, and DIDON—within a unified DON-based architecture.

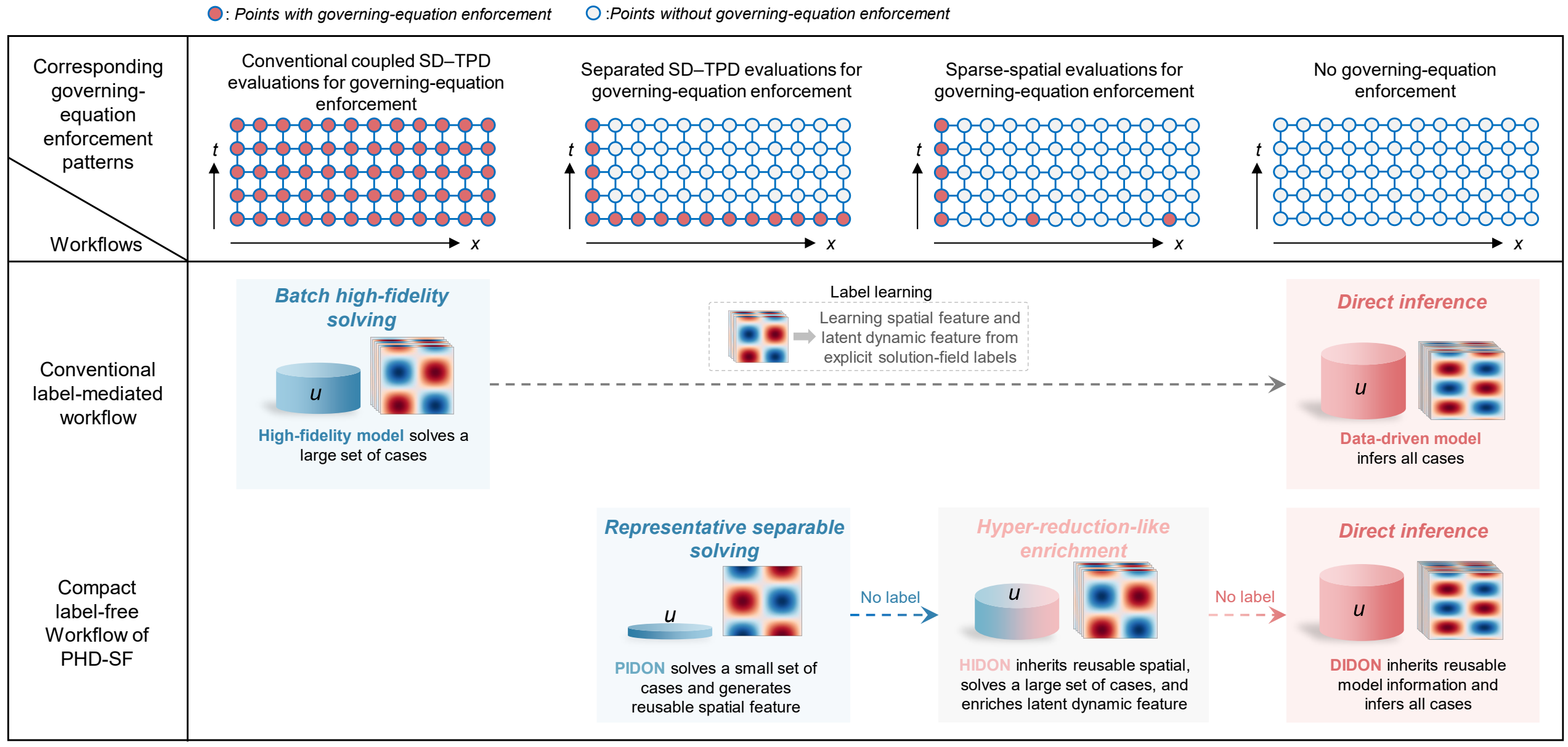

**Fig. 4| Conceptual comparison between the conventional label-mediated workflow and PHD-SF workflow in terms of governing-equation enforcement patterns.** The columns represent different governing-equation enforcement patterns, including conventional coupled SD–TPD evaluations, separated SD–TPD evaluations, sparse-spatial evaluations, and direct inference without governing-equation enforcement. The rows represent different workflows. In the conventional label-mediated workflow, high-fidelity solvers generate solution-field labels through coupled SD–TPD evaluations, after which downstream data-driven models learn from these labels and perform direct inference. In PHD-SF, PIDON operates under separated SD–TPD evaluations, HIDON under sparse-spatial governing-equation enforcement, and DIDON performs direct inference without further governing-equation enforcement.

## 2. Methodology

### 2.1 Problem statement

Set the spatial domain (SD) be $\boldsymbol{x} \in \Omega_x \subset \mathbb{R}^{d_x}$ ,and the time–parameter domain (TPD) be $\boldsymbol{z} = [t, \boldsymbol{\mu}]^T \in \Omega_z \subset \mathbb{R}^{1+d_\mu}$ , where $\boldsymbol{\mu} \in \mathbb{R}^{d_\mu}$ denotes the system parameter vector, including PDE coefficients, source terms, and quantities related to boundary and initial parameters. The unknown state field is $u : \Omega_x \times \Omega_z \to \mathbb{R}^{d_u}$ . We consider a general time-dependent parametric PDE written in operator form as

$$\partial_t u + \mathcal{N}(u, \mathbf{L}u) = 0. \tag{1}$$

Here $\mathbf{L} := [\mathcal{L}_1, \ldots, \mathcal{L}_m]^T$ , $\mathcal{L}_j \in D$ , where $D$ is a dictionary of spatial differential operators (e.g., $\partial_{x_k}$ , $\partial_{x_k x_\ell}$ , $\nabla$ , $\Delta$). The operator $\mathcal{N}$ is a general nonlinear operator including convection, diffusion, reaction, source terms.

### 2.2 PIDON solving and spatial feature learning

PIDON is an accelerated physics-informed neural solver for high-dimensional PDE, designed to efficiently solve a small number of representative parameter cases while simultaneously learning spatial features and differential information of the solution field, as shown in Fig. 1a and Fig. 3a. To achieve linearly scalable computational complexity, PIDON adopts the space–time separation strategy inspired by numerical ROMs. Specifically, PIDON is explicitly decomposed into a time–parameter network (TPNet) and a spatial network (SpatialNet): the former outputs basis coordinate functions (BCFs) to capture the latent time–parameter dynamic features, while the latter outputs basis functions (BFs) to encode the spatial features of the solution manifold. Together, the BCFs and BFs constitute the two separable latent factors of the approximate solution.

To explicitly extract reusable model information, we further introduce a differential-operator learning strategy: transferring the differentiation of the approximate solution to the BCFs and BFs (see Fig. 2b). In this way, once PIDON completes solution of the representative parameter cases, the learned BFs and their corresponding spatial differential operators (SDOs) encode both the spatial features and the associated differential information of the solution field. These quantities can then be inherited by HIDON and DIDON as reusable model information, without requiring an additional supervised learning stage based on PIDON-generated solution labels.

### 2.2.1 SD–TPD separation strategy

To achieve linearly scalable computational complexity, we adopt the classical space–time separation idea from numerical ROMs and refine it into an explicit separation of SD dependence from TPD dependence in the solution representation. Then, PIDON illustrated in Fig. 1a, is explicitly decomposed into a SpatialNet $\mathbf{f}_{\theta_x}:\Omega_x \to \mathbb{R}^{n_b}$ and a TPNet $\mathbf{g}_{\theta_t}:\Omega_z \to \mathbb{R}^{n_b}$. SpatialNet outputs BFs $\mathbf{\Phi}(\boldsymbol{x}) := \mathbf{f}_{\theta_x}(\boldsymbol{x}) = \left(\varphi_1(\boldsymbol{x}),\ldots,\varphi_n(\boldsymbol{x})\right)^T \in \mathbb{R}^{n_b}$, while TPNet outputs BCFs $\mathbf{A}(\boldsymbol{z}) := \mathbf{g}_{\theta_t}(\boldsymbol{z}) = \left(a_1(\boldsymbol{z}),\ldots,a_n(\boldsymbol{z})\right)^T \in \mathbb{R}^{n_b}$, $n_b$ is the number of BFs. The approximate solution of the parametric PDE is then expressed as

$$u(\boldsymbol{x},\boldsymbol{z}) := \mathbf{\Phi}(\boldsymbol{x})^T \mathbf{A}(\boldsymbol{z}) = \sum_{i=1}^{n_b} \varphi_i(\boldsymbol{x}) a_i(\boldsymbol{z}). \tag{2}$$

A natural question is whether such a separable representation is sufficiently expressive for general parametric PDE solutions. The answer is affirmative: the PIDON separable function class is uniformly dense in the space of continuous solutions on compact domains. This implies that any continuous parametric PDE solution can be approximated arbitrarily well by a PIDON representation. The proof combines the classical density of finite sums of separable functions [32, 33] with the universal approximation theorem for neural networks [34-36], and is provided in Appendix S2.

This SD–TPD separation preserves the flexibility of spatial sampling, without being strictly constrained by the sampling rules of all-variable separation frameworks such as proper generalized decomposition (PGD) and separable physics-informed neural networks (SPINNs) [26, 27]. Furthermore, the BFs and BCFs themselves have clear physical meanings, which helps to maintain the interpretability of the model.

### 2.2.2 Differential-operator learning strategy

Based on the separable SD–TPD representation in Equation (2), temporal differentiation of the approximate solution can be transferred to the BCFs, and spatial differentiation of the approximate solution can be transferred to the BFs. Specifically,

$$\begin{aligned}\partial_t u(x,t,\mu) &= \sum_{i=1}^{n}\varphi_i(x)\partial_t a_i(t,\mu) = \mathbf{\Phi}\partial_t\mathbf{A} \\ \mathbf{L}(u)(x,t,\mu) &= \sum_{i=1}^{n}(\mathbf{L}\varphi_i)(x)a_i(t,\mu) = \mathbf{L\Phi A}\end{aligned} \tag{3}$$

Here, SDOs, $\mathbf{L\Phi}$, is defined entrywise by $(\mathbf{L\Phi})_{j,i} = \mathcal{L}_j\varphi_i$. Motivated by Equation (3), we introduce a new differentiation strategy (differential-operator learning strategy) schematically shown in Fig. 2b.

Specifically, AD is applied to the BCFs and BFs with respect to their own variables, thereby transferring the differentiation of the approximate solution to its separable latent factors and explicitly obtaining the corresponding SDOs.

It should be noted that differential-operator learning here does not aim to identify an unknown analytical form of the governing PDE operator, nor is it used in the conventional operator-learning sense of learning a solution operator from input functions or parameters to output solution fields. Rather, it explicitly obtains the differential-operator information induced by the known governing equations on the two separable latent factors. Specifically, temporal differentiation is applied to the BCFs, while the governing spatial differential operators are applied to the learned BFs to obtain the SDOs.

### 2.2.3 PIDON solving

Using the separable SD–TPD representation in Equation (2), and the differential transformation in in Equation (3). The PDE system in Equation (1) can be written abstractly as

$$\partial_t u + \mathcal{N}(u, \mathbf{L}u) = 0 \;\Rightarrow\; \mathbf{\Phi}\partial_t\mathbf{A} + \mathcal{N}(\mathbf{\Phi A}, (\mathbf{L\Phi})\mathbf{A}) = 0. \tag{4}$$

Then, a weighted mean-square-error (MSE) loss

$$L(\theta_x,\theta_z) = \lambda_r L_r + \lambda_b L_b + \lambda_i L_i + \lambda_{orth} L_{orth},$$

can be constructed on the on the spatial collocation set $\mathcal{X}_\mathrm{p} = \{x_i\}_{i=1}^{N_x} \subset \Omega_x$ and the temporal–parametric collocation set $\mathcal{Z}_\mathrm{P} = \{z_i\}_{i=1}^{N_z^\mathrm{P}} \subset \Omega_z^\mathrm{P}$ using $\mathbf{\Phi}$, $\mathbf{L\Phi}$, $\mathbf{A}$, and $\partial_t\mathbf{A}$. where $\{\lambda_\mathrm{r}, L_\mathrm{r}\}$, $\{\lambda_\mathrm{b}, L_\mathrm{b}\}$, $\{\lambda_\mathrm{i}, L_\mathrm{i}\}$, and $\{\lambda_\mathrm{orth}, L_\mathrm{orth}\}$ denote the weights and residual losses associated with the interior PDE, boundary conditions, initial conditions, and orthogonality regularization, respectively. The detailed definitions of these loss terms are provided in Appendix S4.

The role of $L_\mathrm{orth}$ is to improve the orthogonality of $\mathbf{\Phi}$ while maintaining numerical stability, thereby enlarging the effective subspace spanned by $\mathbf{\Phi}$ and enhancing the expressive power and cross-parameter generalization of subsequent HIDON and DIDON. This design echoes classical POD, where orthogonal bases are sought to improve subspace quality. As a flexible penalty, $L_\mathrm{orth}$ provides a controllable trade-off between early-stage optimization cost and the richness of the learned subspace: a stronger orthogonality constraint may increase the initial training cost, but often yields a more representative subspace and thus benefits downstream modelling and generalization. After defining $L(\theta_x, \theta_z)$, we update $\{\theta_x, \theta_z\}$ using Adam (or other optimizers) until convergence, obtaining the trained SpatialNet and the corresponding approximation $\boldsymbol{u}(\boldsymbol{x}, t, \boldsymbol{\mu}; \theta_x^*, \theta_z)$.

### 2.2.4 Computational efficiency and convergence behaviour of PIDON

Owing to the SD–TPD separable architecture, PIDON evaluates the SpatialNet only at $N_x$ distinct spatial collocation points and the TPNet only at $N_z$ distinct temporal–parametric collocation points. Therefore, the required neural-network evaluations and input-based automatic differentiation (AD) operations are performed on $(N_x + N_z)$, rather than with the full tensor-product size $N_x N_z$ as in a conventional PINN. More specifically, because temporal differentiation is applied to the BCFs and spatial differentiation is applied to the BFs, the generation of latent-factor values and their differential information scales as $\mathcal{O}(n_\mathrm{b}(N_x + N_z))$, where $n_\mathrm{b}$ is the number of BFs. By contrast, a conventional PINN evaluates the approximate solution and its derivatives directly at $\mathcal{O}(N_x N_z)$ coupled collocation points. This estimate mainly concerns the dominant costs of network evaluation and input-based AD, which are typically the most expensive parts of high-resolution physics-informed training. Therefore, the SD–TPD separation substantially reduces the practical training cost when $n_\mathrm{b} \ll \min(N_x, N_z)$.

The separated architecture also improves empirical optimization behaviour. In conventional PINNs, spatial and temporal derivatives are imposed on a single network, and high-order differentiation often induces gradient stiffness and hinders training [37]. By contrast, PIDON assigns spatial differentiation to the BFs and temporal differentiation to the BCFs, thereby reducing the differentiation burden on each network and leading to more stable convergence in practice.

### 2.3 HIDON Solving and latent dynamical feature learning

HIDON is an accelerated hybrid solver for a reduced ordinary differential equation (ODE) system, designed to solve a large set of parameter cases at low computational cost while enriching latent dynamical features, as shown in Fig. 1b and Fig. 3b. To reduce the computational cost, HIDON is inspired by the latent-space reduction idea of PROMs and the sparse-element discretization idea of HROMs. On the one hand, by inheriting the BFs, SDOs, and SpatialNet learned by PIDON, HIDON represents both the high-dimensional solution and the associated spatial differential operation in a low-dimensional latent solution space. Specifically, the BFs provide the low-dimensional representation of the solution, while the SDOs encode the spatial differential information required for governing-equation enforcement. In this way, the original high-dimensional PDE is recast into a reduced latent ODE system.

On the other hand, HIDON enforces the governing equations only on a sparse spatial collocation set, producing a hyper-reduction-like computational effect without following a standard HROM construction such as DEIM or ECSW. This sparse enforcement reduces the cost of governing-equation evaluation and enables HIDON to solve a large set of parameter cases at low cost. During this process, the BCFs are further optimized over the temporal–parametric domain, thereby enriching the latent dynamical features encoded in them. These enriched BCFs are then transferred to DIDON as reusable model information, without requiring an additional supervised learning stage based on HIDON-generated solution labels.

#### 2.3.1 Inherit-then-enrich strategy

For many parametric PDEs of practical interest, the solution manifold can often be well approximated by a low-dimensional solution subspace, as suggested by Kolmogorov *n*-width arguments and classical ROM studies [38-40]. In PIDON, such an approximate solution subspace, spanned by the learned BFs, is obtained from a small number of representative parameter cases through physics-informed solving. Although the dominant solution subspace can often be captured from limited representative cases, the latent dynamical features encoded in the BCFs over the TPD generally require richer temporal–parametric sampling to be learned reliably. HIDON therefore adopts an inherit-then-enrich strategy: it inherits the BFs, SDOs, and SpatialNet learned by PIDON, while continuing to train the TPNet over a larger set of parameter cases to enrich the BCFs and the latent dynamical features encoded in them.

### 2.3.2 HIDON solving and latent dynamical feature learning

Building on the inherit–then–enrich strategy, HIDON inherits the $\mathbf{\Phi}$ and $\mathbf{L\Phi}$, learned by PIDON, and substitutes them into the original PDE in Equation (1). The spatial features encoded in $\mathbf{\Phi}$ restrict the high-dimensional solution $\boldsymbol{u}$ to a low-dimensional latent solution space, where $\mathbf{A}$ serves as the corresponding latent state, while $\mathbf{L\Phi}$ provides the spatial differential information required for governing-equation enforcement. Therefore, the only differentiated unknown in HIDON is $\mathbf{A}$ and the original PDE system can be recast into a reduced latent ODE system:

$$\mathbf{\Phi}\frac{d\mathbf{A}}{dt} + \mathcal{N}(\mathbf{\Phi A}, (\mathbf{L\Phi})\mathbf{A}; \boldsymbol{x}, \boldsymbol{z}) = 0, \tag{5}$$

where $\boldsymbol{x} \in \mathcal{X}_{\mathrm{p}} = \{x_i\}_{i=1}^{N_x} \subset \Omega_x$ and $\boldsymbol{z} \in \mathcal{Z}_{\mathrm{H}} = \{z_i\}_{i=1}^{N_z^{\mathrm{H}}} \subset \Omega_z^{\mathrm{H}}$. According to the inherit–then–enrich strategy, HIDON is required to solve and train this ODE system over a large number of parameter cases, i.e. $N_z^{\mathrm{H}} \gg N_z^{\mathrm{P}}$. Compared with the original PDE, the reduced latent ODE system significantly lowers the computational complexity and memory requirement. However, if governing-equation enforcement is still performed on the full high-resolution spatial collocation set $\mathcal{X}_{\mathrm{p}}$, the cost remains proportional to the number of spatial points $N_x$, which becomes a bottleneck for large-scale multi-case training.

To remove this bottleneck, we introduce sparse spatial sampling strategy that follows the same complexity-reduction principle as HROMs. Specifically, the governing equations are enforced only on a sparse yet representative subset of spatial points, so that the computational cost becomes independent of the original spatial resolution. Concretely, HIDON inherits $\mathbf{\Phi}$ and $\mathbf{L\Phi}$ from PIDON and drives the reduced dynamics only on a sparse spatial sampling set $\boldsymbol{x}_\mathrm{s} \in \mathcal{X}_\mathrm{s} = \{x_i\}_{i=1}^{n_\mathrm{s}} \subset \Omega_x$:

$$\mathbf{\Phi}_\mathrm{s} \frac{d\mathbf{A}}{dt} + \mathcal{N}(\mathbf{\Phi}_\mathrm{s}\mathbf{A}, (\mathbf{L\Phi}_\mathrm{s})\mathbf{A}; \boldsymbol{x}_\mathrm{s}, \boldsymbol{z}) = 0, \tag{6}$$

where $\mathbf{\Phi}_\mathrm{s} := \mathbf{f}_{\theta_x}(\boldsymbol{x}_\mathrm{s})$, $\boldsymbol{z} \in \mathcal{Z}_\mathrm{H}$. The sparsity is reflected by $n_\mathrm{s} \ll N_x$, and $\mathbf{\Phi}_\mathrm{s}, \mathbf{L\Phi}_\mathrm{s}$ are precomputed and stored before HIDON training. A weighted MSE loss is then constructed over $\mathcal{X}_\mathrm{s}$ and $\mathcal{Z}_\mathrm{H}$:

$$L(\theta_z) = \lambda_\mathrm{r} L_\mathrm{r} + \lambda_\mathrm{b} L_\mathrm{b} + \lambda_\mathrm{i} L_i, \tag{7}$$

where $\{L_\mathrm{r}, \lambda_\mathrm{r}\}$, $\{L_\mathrm{b}, \lambda_\mathrm{b}\}$, $\{L_\mathrm{i}, \lambda_\mathrm{i}\}$ denote the residual losses are their weights associated with the reduced ODEs, boundary conditions and initial conditions, respectively. Detailed definitions are provided in Appendix S5. During HIDON training, only the parameters of TPNet, $\theta_z$, are updated, while SpatialNet is kept frozen. Minimizing $L(\theta_z)$ via Adam (or other optimisers) yields a converged parameter set $\theta_z^*$ and the corresponding BCFs $\mathbf{A}$. Finally, inserting $\mathbf{A}$ into the separable representation in Equation (2) yields the final approximate solution $\boldsymbol{u}(\boldsymbol{x}, t, \boldsymbol{\mu}; \theta_x, \theta_z)$, enabling efficient large-scale multi-case solution and latent-dynamics learning.

Importantly, the strength of physical constraints in HIDON can be adjusted through the number of sparse spatial sampling points $n_\mathrm{s}$. Therefore, HIDON can vary continuously along the physics-to-data spectrum: using more sparse points strengthens governing-equation enforcement, whereas using fewer points further reduces computational cost and moves the model closer to data-dominated inference.

### 2.3.3 Computational efficiency and closure analysis for HIDON

HIDON inherits the BFs and SDOs learned by PIDON and incorporates hyper-reduction through sparse spatial sampling, so that the original PDE system can be recast into a reduced latent ODE system. This yields a two-level reduction in complexity, both in the solution degrees of freedom and in the cost of governing-equation enforcement. First, the high-dimensional solution is represented in the learned BF space with dimension $n_\mathrm{b}$. Second, the governing equations are enforced only on a sparse spatial set $\mathcal{X}_\mathrm{s} = \{x_i\}_{i=1}^{n_\mathrm{s}}$, rather than on the full spatial collocation set of size $N_x$. As a result, the effective computational cost is mainly determined by $n_{\mathrm{b}}$ and $n_{\mathrm{s}}$. In this sense, HIDON produces a hyper-reduction-like acceleration effect, although it is not a classical HROM.

We next examine the closure property of HIDON. Closure of the resulting reduced latent ODE system depends primarily on whether the sparse governing-equation constraints provide sufficient information to determine the latent state $\mathbf{A}$. For a fixed $(t, \boldsymbol{\mu})$, Let $\mathbf{F}_{t,\mu}(\mathbf{A}) \in \mathbb{R}^{n_{\mathrm{s}}}$ denote the vector of scalar governing-equation constraints evaluated on $\mathcal{X}_{\mathrm{s}}$. A necessary dimension-counting condition for closure is $n_{\mathrm{s}} \geq n_{\mathrm{b}}$. However, this condition alone is not sufficient because the sampled constraints may be linearly dependent or weakly sensitive to some latent directions. A stronger local identifiability condition is that the sparse residual Jacobian $\boldsymbol{J}_{t,\mu} = (\partial \mathbf{F}_{t,\mu} / \partial \mathbf{A})$ has full column rank, so that perturbations in the latent coefficients can be locally distinguished. In this work, we adopt a mildly over-constrained setting together with a simple uniform-sampling baseline for point selection. These choices provide a practical balance between robustness and efficiency. Although $n_{\mathrm{s}}$ and $\mathcal{X}_{\mathrm{s}}$ are both important and interesting, a full exploration is beyond the scope of this manuscript, and a detailed analysis is deferred to Appendix S3.

### 2.4 Direct inference via DIDON

DIDON is constructed by directly inheriting the trained SpatialNet from PIDON and the enriched TPNet from HIDON. It performs online prediction for parameter cases within the enriched temporal–parametric domain $\mathcal{Z}_{\mathrm{D}} = \{z_i\}_{i=1}^{N_z^{\mathrm{D}}} \subset \Omega_z^{\mathrm{D}}$. Since HIDON has already enriched the latent dynamical features over the target temporal–parametric domain, this work focuses on DIDON inference within this enriched domain, rather than on extrapolation beyond it. Therefore, we set $\Omega_z^{\mathrm{D}} = \Omega_z^{\mathrm{H}}$ in the numerical experiments. Under this setting, DIDON attains accuracy comparable to HIDON while requiring only negligible online computational cost.

### 2.5 PHD-SF

PHD-SF is a compact physics-to-data spectrum framework for accelerating parametric time-dependent PDEs. For clarity, it is described as consisting of three components: PIDON, HIDON, and DIDON. More fundamentally, these components are three operating modes of a unified DON-based architecture at different stages of the physics-to-data spectrum. A more detailed interpretation is provided in Appendix S1. Specifically, PHD-SF switches among different operating modes through parameter reuse, staged training and freezing of subnetworks, thereby enabling the cross-stage transfer of reusable model information among PIDON, HIDON, and DIDON.

Fig. 4 conceptually compares PHD-SF workflow with a conventional label-mediated workflow from high-fidelity solvers to data-driven models. The figure highlights two key differences. First, the carrier connecting different modeling stages changes from explicit solution-field labels to reusable model information. Second, the governing-equation enforcement pattern changes from coupled space–time sampling followed by direct inference in the conventional workflow, to separated SD–TPD sampling in PIDON, sparse spatial sampling in HIDON, and direct inference in DIDON.

From the perspective of information connection, PHD-SF is a compact label-free generation–inheritance–enrichment workflow of reusable model information. As shown in Fig. 4, the conventional label-mediated workflow usually requires high-fidelity solvers to generate explicit solution labels, which are then used by downstream data-driven models to learn spatial features and latent dynamical features. In contrast, the PHD-SF workflow connects its three operating modes, namely PIDON, HIDON, and DIDON, through reusable model information generated, inherited, and enriched within a unified DON-based architecture. The term "label-free" has two implications. Externally, PHD-SF does not require precomputed high-fidelity solution snapshots as training labels. Internally, there is no explicit solution-field label collection or label-based learning among PIDON, HIDON, and DIDON.

From the perspective of computational complexity, PHD-SF substantially reduces the total end-to-end computational cost through a compact "representative separable solving–hyper-reduction-like enrichment–direct inference (solve–enrich–infer)" acceleration path. As shown in Fig. 4, the conventional workflow follows the path of "batch high-fidelity solving–solution-label-based training–direct inference." Although this workflow can provide fast online inference after training, it usually requires high-fidelity solvers to precompute a large set of parameter cases for dataset construction. This offline data-generation stage introduces a considerable computational burden and remains a key limitation of data-driven accelerators. In comparison, the PHD-SF follows a solve–enrich–infer acceleration path. In PIDON, the SD–TPD separable representation reduces the computational complexity of physics-informed solving, while only a small number of representative parameter cases are solved, avoiding the construction of large-scale high-fidelity solution datasets. In HIDON, the inherited spatial features and differential information are reused, and governing-equation constraints are enforced only on sparse spatial collocation points, leading to a hyper-reduction-like acceleration effect and significantly reducing the cost of latent-dynamics enrichment over multiple parameter cases. Finally, DIDON directly reuses the model information inherited from PIDON and HIDON for fast inference. In this way, PHD-SF improves the end-to-end computational efficiency of many-query parametric time-dependent PDE problems while retaining the role of physical constraints.

In summary, PHD-SF has three key characteristics. First, at the framework level, it organizes the generation, inheritance, and enrichment of reusable model information within a unified DON-based architecture. Second, at the individual-mode level, solving and reusable-information learning are performed simultaneously, so that reusable model information is formed during the solving process itself. Third, at the spectrum level, PHD-SF establishes a progressive acceleration path from strongly physics-constrained solving, to hybrid latent-dynamics enrichment, and finally to fast direct inference. Along this path, explicit physical constraints gradually weaken, whereas inherited reusable model information becomes increasingly important, providing a unified framework for balancing physical consistency, cross-parameter reusability, and inference efficiency.

Algorithm 1 summarizes the overall computational pipeline of PHD-SF, including PIDON, HIDON, and DIDON. The detailed procedures of each stage are listed below.

**Algorithm 1**. PHD-SF pipeline

**Input**: PDE operator $\mathcal{N}$ ;SD collocation set $\mathcal{X}_{\mathrm{p}}$; PIDON solving TPD collocation set $\mathcal{Z}_{\mathrm{P}}$; HIDON enrichment TPD collocation set $\mathcal{Z}_{\mathrm{H}}$ DIDON inference TPD collocation set $\mathcal{Z}_{\mathrm{D}}$; sparse SD set $\mathcal{X}_{\mathrm{s}} \subset \mathcal{X}_{\mathrm{p}}$.

**Output**: trained SpatialNet $\mathbf{\Phi}(\boldsymbol{x};\theta_x^*)$; TPNet $\mathbf{A}(t,\boldsymbol{\mu};\theta_z^*)$; PIDON/HIDON solvers and DIDON inference model.

**1. PIDON: solve PDE system & learn spatial features**

1.1 For $\boldsymbol{x} \in \mathcal{X}_{\mathrm{p}}$, evaluate BFs $\mathbf{\Phi}(\boldsymbol{x};\theta_x)$ and compute SDOs $\mathbf{L\Phi}$ via AD.

1.2 For $(t,\boldsymbol{\mu}) \in \mathcal{Z}_{\mathrm{P}}$, evaluate BCFs $\mathbf{A}(t,\boldsymbol{\mu};\theta_z)$ and compute $\partial_t \mathbf{A}$ via AD.

1.3 Assemble residuals and compute $L(\theta_x, \theta_z)$.

1.4 Update $\{\theta_x, \theta_z\} \leftarrow \text{Adam}(\nabla_{\theta_x,\theta_z} L)$ until convergence, obtaining $\theta_x^*$, $\mathbf{\Phi}(\boldsymbol{x};\theta_x^*)$, $\mathbf{L\Phi}$ and $\boldsymbol{u}$.

**2. Precompute for HIDON**

2.1 For $\boldsymbol{x}_\text{s} \in \mathcal{X}_\text{s}$, calculate $\mathbf{\Phi}_\text{s} = \mathbf{\Phi}(\boldsymbol{x}_\text{s};\theta_x^*)$ and $\mathbf{L\Phi}_\text{s}$.

**3. HIDON: solve reduced latent ODE system & learn latent dynamical features**

3.1 For $(t, \boldsymbol{\mu}) \in \mathcal{Z}_\text{H}$, evaluate BCFs $\mathbf{A}(t, \boldsymbol{\mu}; \theta_z)$ and compute $\partial_t \mathbf{A}$ via AD.

3.2 Assemble residuals and compute $L(\theta_z)$.

3.3 Update $\theta_z \leftarrow \text{Adam}(\nabla_{\theta_z} L)$ until convergence, obtaining $\theta_z^*$, $\mathbf{A}(t, \boldsymbol{\mu}; \theta_z^*)$ and $\boldsymbol{u}$.

**4. DIDON: direct inference**.

4.1 For $(t, \boldsymbol{\mu}) \in \mathcal{Z}_\text{D}$, direct infer $\boldsymbol{u}(\boldsymbol{x}, t, \boldsymbol{\mu}; \theta_x^*, \theta_z^*) = \mathbf{\Phi}(\boldsymbol{x};\theta_x^*)\mathbf{A}(t, \boldsymbol{\mu}; \theta_z^*)$.

## 3. Results

We evaluate PHD-SF on four benchmark time-dependent PDE problems using deliberately sparse parameter configurations, as summarized in Table 1. These configurations are referred to as one-shot-like because only one or two representative parameter cases are used in the initial PIDON solving stage. Since PHD-SF follows this compact label-free solve–enrich–infer acceleration path rather than the conventional label-mediated computational pattern, we do not adopt a global training–test split. Instead, the parametric sets are named according to their roles in the three operating modes: the PIDON solving set $D_{\mathrm{P}}$, the HIDON enrichment set $D_{\mathrm{H}}$, and the DIDON inference set $D_{\mathrm{D}}$. The target parameter domain covered by the HIDON enrichment set and the DIDON inference set is denoted as $\Omega_{\mu}^{\mathrm{T}}$.

Detailed network architectures and hyperparameters are provided in Appendix S6, and the detailed collocation-set sizes used in all benchmarks are provided in Appendix S7. To ensure a fair comparison, all models are trained under an equivalent full-batch optimization setting. For PINN, a literal full-batch forward/backward pass may exceed GPU memory limits because of the large number of coupled collocation points. We therefore use gradient accumulation to emulate full-batch updates under GPU memory constraints. This implementation may increase the wall-clock training time of PINN, but it ensures that all models are compared under the same full-batch optimization protocol. A detailed explanation is provided in Appendix S8. All training and inference procedures are implemented in PyTorch and executed on an NVIDIA GeForce RTX 5070 Ti GPU (16 GB memory).

**Table 1. Summary of benchmarks for assessing the computational task and performance of each model.** The reported solution error corresponds to the relative $L_2$ error averaged over all examples in the test dataset.

| Governing law | Model | Computational task | Run-time (s) | Error | Memory (MB) |
|---|---|---|---|---|---|
| **Diffusion reaction:** $u_t = \lambda u_{xx} + ru$ $u(x,0) = u_0(x;\alpha)$ **with** $\boldsymbol{\mu} = (r,\alpha)$ | PINN | $D_{\mathrm{P}} = \{D(\boldsymbol{\mu}_1)\},\ \boldsymbol{\mu}_1 = (3,1)$ | 444,120 | | 7,346 |
| | PIDON | $D_{\mathrm{P}} = \{D(\boldsymbol{\mu}_1)\},\ \boldsymbol{\mu}_1 = (3,1)$ | 49,228 | | 7,434 |
| | HIDON | $D_{\mathrm{H}} = \{D(\boldsymbol{\mu}_j)\}_{j=1}^{121},\ \boldsymbol{\mu}_j \in [1,3]\times[-1,1]$ | 1,262 | 0.35±0.10% | 3978 |
| | DIDON | $D_{\mathrm{D}} = \{D(\boldsymbol{\mu}_j)\}_{j=1}^{2500},\ \boldsymbol{\mu}_j \in [1,3]\times[-1,1]$ | 0.22 | 0.36±0.10% | 1,984 |
| **Wave:** $u_{tt} = pu_{xx}$ $u(x,0) = u_0(x;\theta)$ | PINN | $D_{\mathrm{P}} = \{D(\boldsymbol{\mu}_1)\},\ \boldsymbol{\mu}_1 = (1,0)$ | 686,160 | | 6,860 |
| | PIDON | $D_{\mathrm{P}} = \{D(\boldsymbol{\mu}_1)\},\ \boldsymbol{\mu}_1 = (1,0)$ | 8,310 | | 7,434 |
| | HIDON | $D_{\mathrm{H}} = \{D(\boldsymbol{\mu}_j)\}_{j=1}^{77},\ \boldsymbol{\mu}_j \in [0.5,3]\times[0,0.75\pi]$ | 3,351 | 0.20±0.08% | 9,410 |
| **with** $\boldsymbol{\mu}=(p,\theta)$ | DIDON | $D_{\mathrm{D}}=\{D(\boldsymbol{\mu}_j)\}_{j=1}^{2500},\ \boldsymbol{\mu}_j\in[0.5,3]\times[0,0.75\pi]$ | 0.53 | 0.18±0.07% | 3,892 |
| **Burger:** $u_t=uu_x-vu_{xx}$ **with** $\boldsymbol{\mu}=(v)$ | PINN | $D_{\mathrm{P}}=\{D(0.01),D(0.1)\}$ | 296,000 | | 11,154 |
| | PIDON | $D_{\mathrm{P}}=\{D(0.01),D(0.1)\}$ | 8,580 | | 9,930 |
| | HIDON | $D_{\mathrm{H}}=\{D(\boldsymbol{\mu}_j)\}_{j=1}^{48},\ \boldsymbol{\mu}_j\in[0.08,0.5]$ | 2,900 | 0.18%±0.26% | 2,400 |
| | DIDON | $D_{\mathrm{D}}=\{D(\boldsymbol{\mu}_j)\}_{j=1}^{250},\ \boldsymbol{\mu}_j\in[0.08,0.5]$ | 0.03 | 0.14±0.18% | 138 |
| **Heat transfer:** $u_t=\alpha(u_{xx}+u_{yy})$ $u(x,1,t)=u(x,t;\theta)$ **with** $\boldsymbol{\mu}=(\alpha,\theta)$ | PINN | $D=\{D(\boldsymbol{\mu}_1)\},\ \boldsymbol{\mu}_1=(1,1)$ | 1,034,780 | | 7,010 |
| | PIDON | $D_{\mathrm{train}}=\{D(\boldsymbol{\mu}_1)\},\ \boldsymbol{\mu}_1=(1,1)$ | 24,962 | | 9,600 |
| | HIDON | $D_{\mathrm{test}}=\{D(\boldsymbol{\mu}_j)\}_{j=1}^{108},\ \boldsymbol{\mu}_j\in[0.75,3]\times[0.3,5]$ | 10,534 | 0.6±0.32% | 13,000 |
| | DIDON | $D_{\mathrm{test}}=\{D(\boldsymbol{\mu}_j)\}_{j=1}^{1500},\ \boldsymbol{\mu}_j\in[0.75,3]\times[0.3,5]$ | 0.18 | 0.62±0.41% | 2,770 |

**Table 2. PIDON and HIDON architectures for each benchmark employed in this work (unless otherwise stated).**

| Governing law | TPNet width | TPNet output width | TPNet depth | SpacialNet width | SpacialNet Output width | SpacialNet depth | Activation function |
|---|---|---|---|---|---|---|---|
| **Diffusion reaction** | 50 | 12 | 5 | 50 | 12 | 5 | Tanh |
| **Wave** | 50 | 12 | 5 | 50 | 12 | 5 | Tanh |
| **Burger** | 50 | 12 | 5 | 50 | 12 | 5 | Tanh |
| **Heat transfer** | 128 | 16 | 6 | 128 | 16 | 6 | Tanh |

### 3.1 Diffusion-reaction equation

We begin with a classical nonlinear diffusion–reaction equation:

$$\begin{aligned}&\text{PDE: } u_t(x,t)=\lambda u_{xx}(x,t)+ru(x,t),\quad x\in[0,1],t\in[0,1],\\&\text{BC: } u(0,t)=u(1,t),u_x(0,t)=u_x(1,t),\quad t\in[0,1],\\&\text{IC: } u(x,0)=\alpha\sin(2\pi x)+\beta\cos(2\pi x),\quad x\in[0,1].\end{aligned}$$

Here, $\alpha$ controls the initial condition (IC) amplitude, and $r$ modulates the reaction strength; together they form the system parameter vector $\boldsymbol{\mu}=(r,\alpha)$. $\lambda$ and $\beta$ are fixed constants.

#### 3.1.1 Solving process

a) PIDON. The weighted loss $L(\theta_x,\theta_z)$ for diffusion–reaction equation reads

$$L(\theta_x,\theta_z)=\lambda_{\mathrm{r}}L_{\mathrm{r}}+\lambda_{\mathrm{b}}L_{\mathrm{b}}+\lambda_{\mathrm{i}}L_{\mathrm{i}}+\lambda_{\mathrm{orth}}L_{\mathrm{orth}}$$

with

$$\begin{aligned}L_{\mathrm{r}}&=\frac{1}{N_xN_Z}\left\|\mathbf{\Phi}(x)\partial_t\mathbf{A}(t,\boldsymbol{\mu})-\lambda\partial_{xx}\mathbf{\Phi}(x)\mathbf{A}(t,\boldsymbol{\mu})-r\mathbf{\Phi}(x)\mathbf{A}(t,\boldsymbol{\mu}))\right\|_F^2,\\L_{\mathrm{b}}&=\frac{1}{N_Z}\left\|(\mathbf{\Phi}(0)-\mathbf{\Phi}(1))\mathbf{A}(t,\boldsymbol{\mu})\right\|_F^2+\frac{1}{N_Z}\left\|(\partial_x\mathbf{\Phi}(0)-\partial_x\mathbf{\Phi}(1))\mathbf{A}(t,\boldsymbol{\mu})\right\|_F^2,\\L_{\mathrm{i}}&=\frac{1}{N_xN_\mu}\left\|\mathbf{\Phi}(x)\mathbf{A}(0,\boldsymbol{\mu})-\alpha\sin(2\pi x)-\beta\cos(2\pi x)\right\|_F^2.\\L_{\mathrm{orth}}&=\max\left(0,\frac{1}{n_{\mathrm{b}}^2}\left\|\mathbf{G}-\mathbf{I}_d\right\|_F^2-\gamma_{\mathrm{orth}}\right),\end{aligned}$$

and $\lambda_{\mathrm{r}}$, $\lambda_{\mathrm{b}}$, $\lambda_{\mathrm{i}}$ and $\lambda_{\mathrm{orth}}$ are 1, 5, 100 and 0.1, respectively. After defining $L(\theta_x,\theta_z)$, we update $\{\theta_x,\theta_z\}$ using Adam until convergence, obtaining the trained $\mathbf{\Phi}$ and the corresponding approximation $\boldsymbol{u}(\boldsymbol{x},t,\boldsymbol{\mu};\theta_x,\theta_z)$.

b) HIDON. We precompute and store the sparse basis evaluations $\mathbf{\Phi}_{\mathrm{s}}$ and their associated spatial differential operator $\mathbf{L\Phi}_{\mathrm{s}} = \{\partial_x \mathbf{\Phi}_{\mathrm{s}}, \partial_{xx} \mathbf{\Phi}_{\mathrm{s}}\}$ on the sparse spatial sampling points $x_{\mathrm{s}} \in \mathcal{X}_{\mathrm{s}} = \{x_i\}_{i=1}^{n_{\mathrm{s}}} \subset \Omega_x$. The weighted loss $L(\theta_z)$ for diffusion–reaction equation is

$$L(\theta_z) = \lambda_{\mathrm{r}} L_{\mathrm{r}} + \lambda_{\mathrm{b}} L_{\mathrm{b}} + \lambda_{\mathrm{i}} L_{\mathrm{i}}$$

with

$$\begin{aligned}
L_{\mathrm{r}} &= \frac{1}{n_{\mathrm{s}} N_Z} \left\| \mathbf{\Phi}_{\mathrm{s}} \frac{d\mathbf{A}}{dt} - \lambda \partial_{xx} \mathbf{\Phi}_{\mathrm{s}} \mathbf{A} - r \mathbf{\Phi}_{\mathrm{s}} \mathbf{A} \right\|_F^2, \\
L_{\mathrm{b}} &= \frac{1}{N_Z} \left\| (\mathbf{\Phi}_{\mathrm{s}}(0) - \mathbf{\Phi}_{\mathrm{s}}(1)) \mathbf{A}(t, \boldsymbol{\mu}) \right\|_F^2 + \frac{1}{N_Z} \left\| (\partial_x \mathbf{\Phi}_{\mathrm{s}}(0) - \partial_x \mathbf{\Phi}_{\mathrm{s}}(1)) \mathbf{A}(t, \boldsymbol{\mu}) \right\|_F^2, \\
L_{\mathrm{i}} &= \frac{1}{n_{\mathrm{s}} N_\mu} \left\| \mathbf{\Phi}_{\mathrm{s}} \mathbf{A}(0, \boldsymbol{\mu}) - \alpha \sin(2\pi x_{\mathrm{s}}) - \beta \cos(2\pi x_{\mathrm{s}}) \right\|_F^2,
\end{aligned}$$

and $\lambda_{\mathrm{r}}$, $\lambda_{\mathrm{b}}$ and $\lambda_{\mathrm{i}}$ are 1, 5 and 100, respectively. It is worth noting that boundary constraints are enforced only on the sparse boundary subset $\partial\Omega_x \cap \mathcal{X}_{\mathrm{s}}$. For the periodic boundary conditions $\partial\Omega = \{0,1\}$, we construct $\mathcal{X}_{\mathrm{s}}$ to explicitly include the endpoints, so that boundary constraints are well-defined. After defining $L(\theta_z)$, we update $\theta_z$ using Adam until convergence, obtaining the trained $\mathbf{A}$ and the corresponding approximate solutions.

c) DIDON. DIDON is constructed by directly inheriting the trained SpatialNet and TPNet from PIDON and HIDON. DIDON is then used for direct solution inference.

### 3.1.2 Computational task and preliminary results

To stress-test the cross-parameter capability of PHD-SF, we adopt a deliberately sparse solve–enrich–infer configuration. PIDON is first solved on one representative parameter case, $D_{\mathrm{P}} = \{D(\boldsymbol{\mu}_1)\}$. HIDON then enriches the latent dynamics on 121 uniformly sampled parameter cases in $\Omega_\mu^{\mathrm{T}} = [1,3] \times [-1,1]$, denoted as the HIDON enrichment set $D_{\mathrm{H}}$. Finally, DIDON performs direct inference on 2500 parameter cases over the same parameter domain, denoted as the inference set $D_{\mathrm{D}}$. The high-resolution spatial and temporal collocation set uses $N_x = 40{,}000$, $N_t = 1{,}000$, with uniform sampling.

This deliberately stringent configuration is not intended to imply that PHD-SF is restricted to single-case or few-case physics-informed solving. Instead, it is used to assess whether HIDON can exploit the BFs and SDOs to enrich the latent dynamics over a broader parameter domain. This setting therefore evaluates the cross-parameter enrichment capability of HIDON, the residual-free direct inference capability of DIDON, and the overall effectiveness of PHD-SF for cross-parameter solution and inference tasks. Specifically, if the solution field obtained by PIDON from a single parameter case were simply treated as externally labeled data for training a conventional data-driven DL-AM, such single-case supervision would generally be insufficient to learn a reliable cross-parameter solution mapping [17]. In PHD-SF, the representative case instead provides reusable spatial features and differential information, which are inherited and enriched under physical constraints in HIDON.

Fig. 5(a–d) visualize PIDON solving set $D_{\mathrm{P}}$, HIDON enrichment set $D_{\mathrm{H}}$, DIDON inference set $D_{\mathrm{D}}$, and the relative $L_2$ error fields of HIDON and DIDON across $D_{\mathrm{H}}$ and $D_{\mathrm{D}}$. In this numerical experiment, the relative $L_2$ error of HIDON and DIDON reach 0.35%±0.1% and 0.36%±0.1%, respectively, with DIDON exhibiting uniformly low absolute error across representative test parameter cases (see Fig. 7), demonstrating the strong cross-parameter inference capability of PHD-SF within the enriched parameter domain.

### 3.1.3 Computational efficiency and accuracy analysis

Since PINN and PIDON both serve as physics-informed PDE solvers, we first compare these two models. Under comparable memory usage, PIDON converges slightly faster in iteration count and reduces wall-clock time by about two orders of magnitude relative to PINN (see Fig. 5e-f), while achieving one order of magnitude better accuracy under nearly identical computational budgets (see Fig. 5h).

HIDON further improves the efficiency of cross-parameter solving. When enriching the latent dynamics over 121 parameter cases simultaneously, HIDON requires substantially less memory than PIDON and its wall-clock time is only about 0.3× that of the PIDON solving stage (Fig. 5e-f). After this enrichment stage, DIDON performs direct inference over 2500 parameter cases with only an additional 0.22 s. Viewed as an integrated solve–enrich–infer path, PIDON+HIDON+DIDON enables direct inference over 2500 parameter cases with a total wall-clock time that is still more than two orders of magnitude lower than that required by PINN to solve a single parameter case. In contrast, the conventional label-mediated workflow illustrated in Fig. 4 would require labeled solution data for many parameter cases before surrogate training; if a PINN-type solver were used to generate labels for the 121 enrichment cases considered here, the offline data-generation cost alone would exceed the total PHD-SF cost by several orders of magnitude.

Transitioning from PIDON to HIDON leads to a moderate loss of accuracy but reduces memory and time costs by roughly three orders of magnitude, whereas the transition from HIDON to DIDON preserves nearly the same accuracy while further reducing memory and time costs by about one and five orders of magnitude, respectively (Fig. 5h).

Finally, Fig. 5i summarizes the overall performance along four axes: physics information, data information, accuracy, and solving speed. In this diagram, physics information represents the strength of explicit governing-equation enforcement, while data information denotes the inherited reusable model information. Along the PIDON–HIDON–DIDON progression, explicit governing-equation enforcement gradually weakens and inherited reusable model information becomes increasingly dominant. This transition substantially improves solving speed with only moderate accuracy degradation, thereby broadening the achievable performance envelope of PHD-SF relative to PINN

#### 3.1.4 BFs analysis

We further examine the influence of the number of BFs, $n_{\mathrm{b}}$, on PHD-SF. As shown in Fig. 6a–c, PIDON exhibits only weak sensitivity to $n_{\mathrm{b}}$, whereas its time and memory costs increase approximately linearly, consistent with the complexity analysis. The cost of SDO evaluation can be further reduced through optimized GPU parallelization. For HIDON and DIDON, the error decreases rapidly at small and then saturates, reflecting the classical behavior of low-dimensional manifold approximation, as discussed in Appendix S9.

A comparison between Fig. 6d and Fig. 6f reveals a clear difference between the PIDON-learned BFs latent space and the POD latent space. PIDON produces diverse and information-rich spatial structures, whereas POD is dominated by a single leading mode and the remaining modes contribute limited additional information. This difference is further quantified by the residual energy (RE) across parameter cases, as shown in Fig. 6(g), with the definition provided in Appendix S10. Except for the PIDON solving case, PIDON achieves RE values that are 3–4 orders of magnitude lower than those of POD in $\Omega_{\mu}^{\mathrm{T}}$, indicating substantially better coverage of the solution manifold and. This improved latent-space coverage explains the strong cross-parameter performance of HIDON and DIDON. Beyond this quantitative comparison, PIDON-learned BFs offer four advantages over POD modes:

a) *Potentially improved latent-space expressiveness.* In this benchmark, the learned BFs span a richer latent space with better cross-parameter coverage.

b) *Smoothness and differentiability.* They are spatially differentiable, and naturally associated with SDOs that encode derivative information, as shown in Fig. 6(e).

c) *Absence of POD-based spectral truncation.* They avoid the truncation error introduced by singular-value truncation in POD (quantified by ΔRE_2).

d) *Transferability.* They can be efficiently adapted under parameter-domain shifts through warm-start transfer.

These properties highlight the expressiveness, smoothness and differentiability, and adaptability of PIDON-learned BFs and SDOs, which are central to the strong generalization performance of PHD-SF and highlight their potential to further advance DL-AMs.

### 3.1.5 Spatial-collocation scaling analysis

We further investigate the influence of the number of spatial collocation points, $N_x$, on accuracy and computational cost (Fig. 6h–j). In this test, $N_\mu$=1 and $N_t$=1000 are fixed, and increasing $N_x$ directly increases the scale of the full spatial collocation set. The errors of PINN and PIDON generally decrease with increasing $N_x$, whereas HIDON and DIDON remain nearly insensitive to $N_x$. Compared with PINN, PIDON shows more favourable cost scaling with respect to $N_x$, supporting the efficiency gain introduced by SD–TPD separation. The run-time and memory consumption of HIDON and DIDON are almost unchanged as $N_x$ increases, indicating that their dominant cost is decoupled from the full spatial sampling scale. This behavior demonstrates the spatial-scale-independent acceleration capability enabled by sparse spatial sampling in HIDON and direct inference in DIDON.

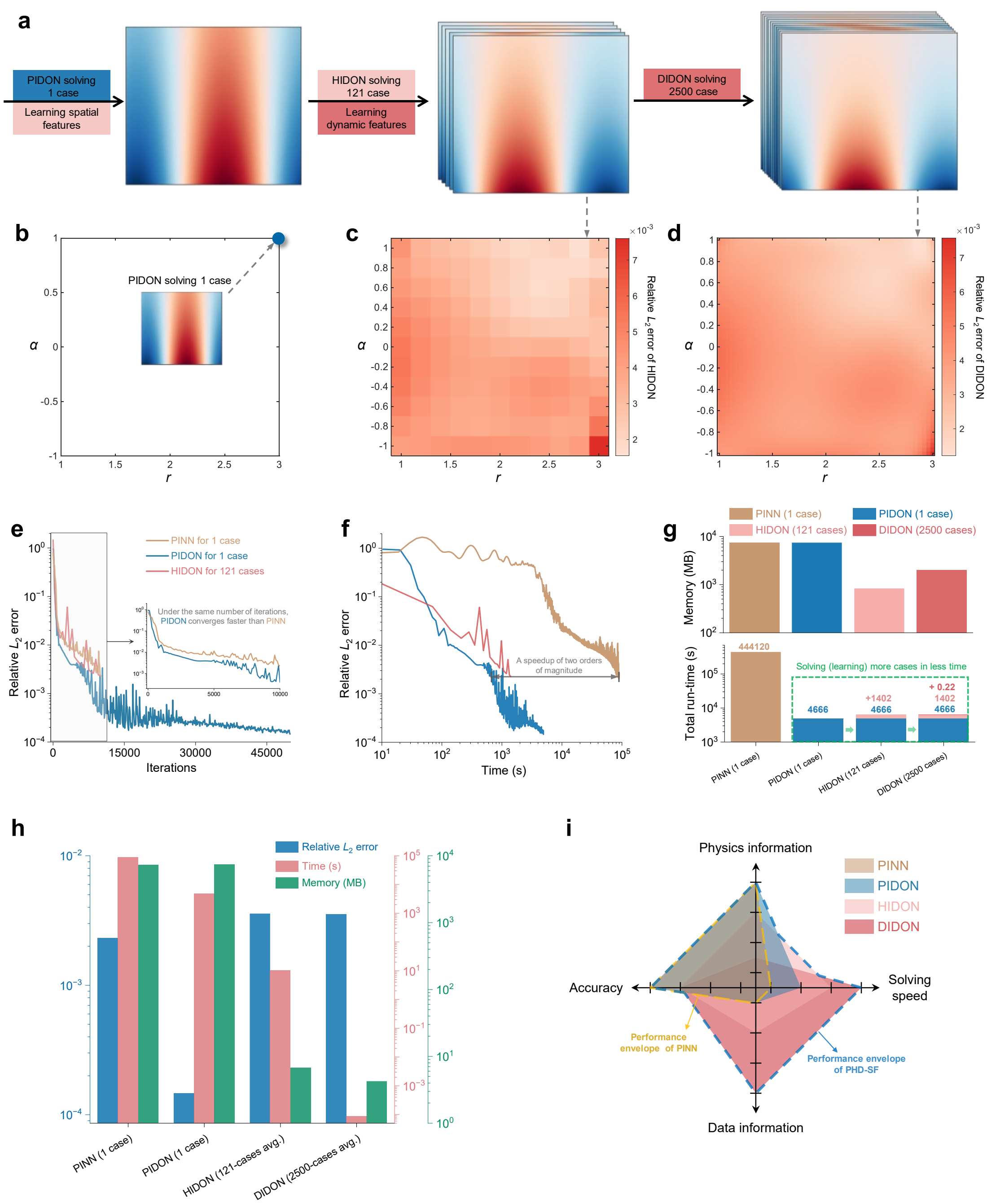

**Fig. 5 | Results for the diffusion–reaction equation. a,** PHD-SF workflow: PIDON is solved on one representative parameter case while learning apatial features; HIDON then solves 121 parameter cases while enriching the latent dynamic features; DIDON finally performs direct inference on 2,500 parameter cases. **b,** PIDON solving case and the target parameter domain. **c–d,** Relative L2 error landscapes of HIDON and DIDON. **e–f,** Convergence histories of conventional PINN, PIDON and HIDON, reported as relative L2 error versus iteration count and wall-clock time under their respective tasks. **g,** GPU memory footprint and total run-time for different models. **h,** Per-case averages of memory footprint, wall-clock time and relative L2 error for PINN, PIDON, HIDON and DIDON; for HIDON and DIDON, the per-case cost is computed as the total cost divided by

the number of solved cases ("n-cases average"). **i,** a qualitative characterization of model performance from physics-information, data-information perspectives, accuracy and solving speed perspectives.

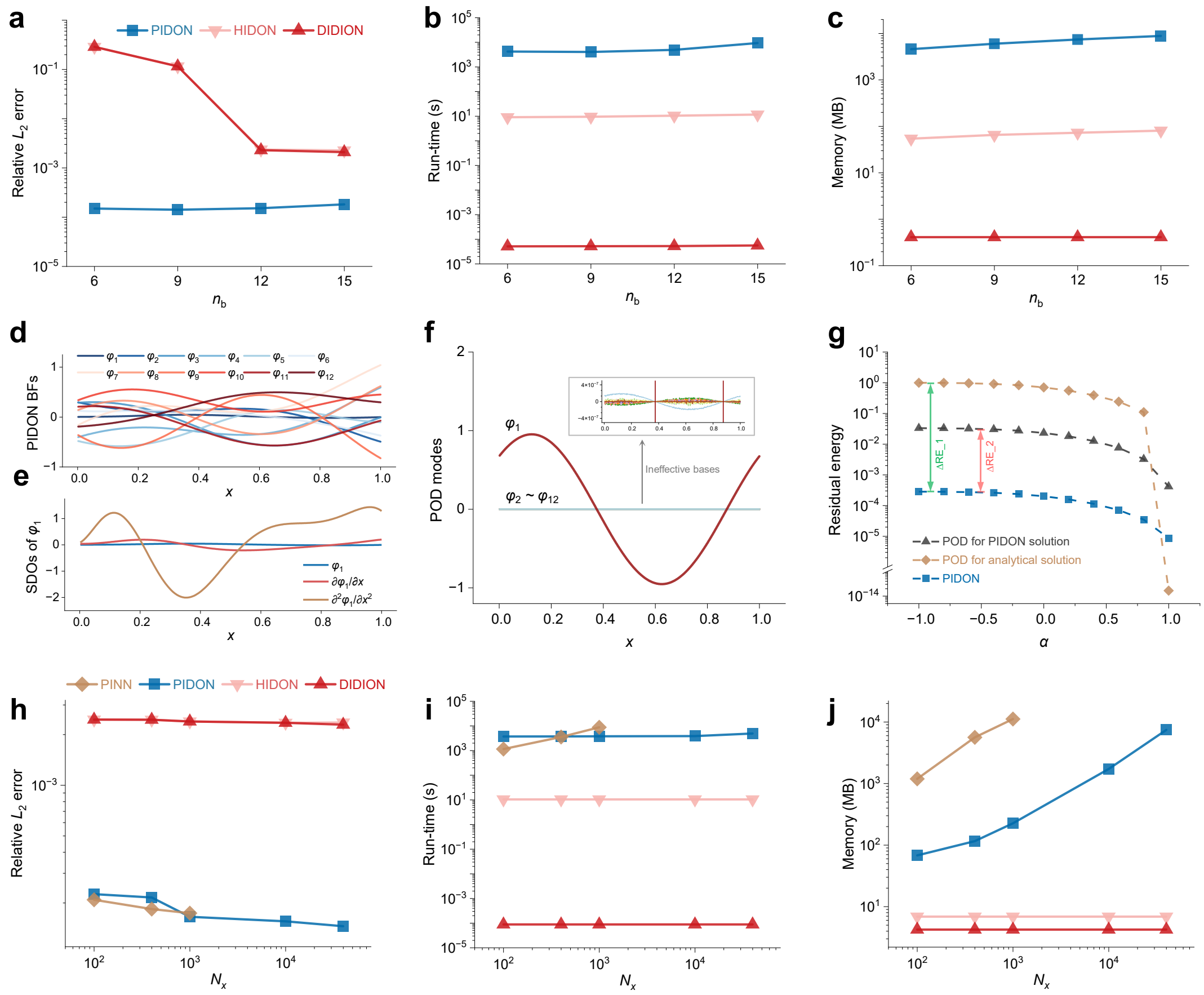

**Fig. 6| BFs and collocation-size analysis. a–c,** Effects of the number of BFs $n_b$ on the relative $L_2$ error, wall-clock time, and GPU memory for different models. **d,** BFs learned by PIDON. **e,** SDOs associated with one of PIDON-learned BFs. **f,** The first 12 normalized POD modes extracted from analytical solutions. **g,** Residual energy (RE) comparison between the subspaces induced by PIDON and POD across parameter cases. **h–j,** Effects of the spatial collocation point, $N_x$, on the relative $L_2$ error, wall-clock time, and GPU memory for different models.

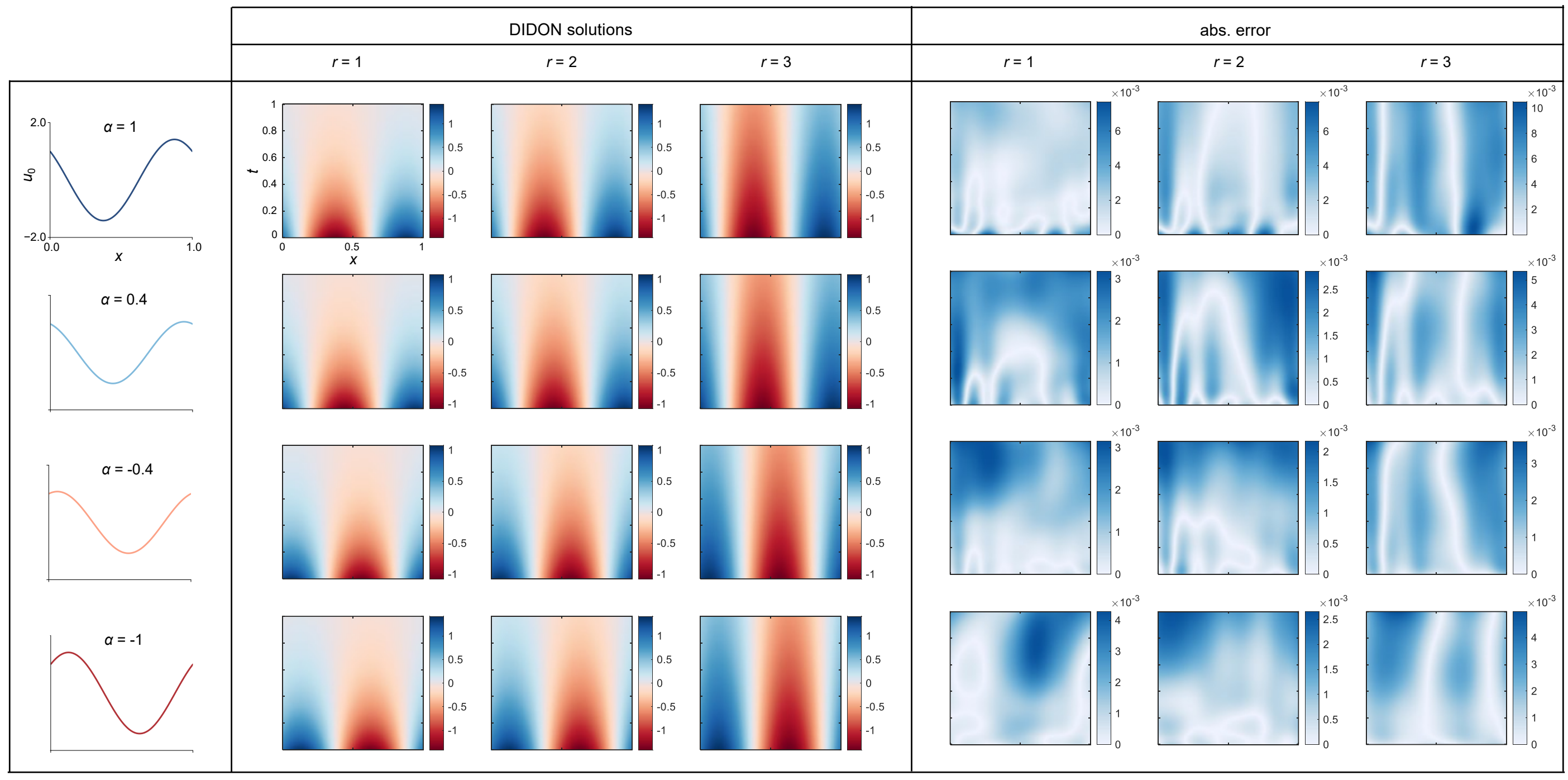

**Fig. 7 | DIDON predicted solution fields and the corresponding absolute error fields for representative parameter cases.**

### 3.2 Wave equation

We next consider a classical wave equation:

$$\begin{aligned}
&\text{PDE: } u_{tt}(x,t) = p u_{xx}(x,t), && x \in [0,1], t \in [0,1],\\
&\text{BC: } u(0,t) = u(1,t), u_x(0,t) = u_x(1,t), && t \in [0,1],\\
&\text{IC: } u(x,0) = A\sin(2\pi x)\cos(\theta), && x \in [0,1],\\
&\quad u_t(x,0) = -A\omega\sin(2\pi x)\sin(\theta), && x \in [0,1],
\end{aligned}$$

here $p = (\omega/2\pi)^2$ controls the PDE coefficient (wave speed) and $\theta$ affects the IC; together they form the system parameter vector $\boldsymbol{\mu}(p,\theta)$. $A$ is fixed constant.

#### 3.2.1 Solving process

a) PIDON. The weighted loss $L(\theta_x,\theta_z)$ for wave equation is defined as

$$L(\theta_x,\theta_z) = \lambda_r L_r + \lambda_b L_b + \lambda_i L_i + \lambda_{orth} L_{orth}$$

with

$$
\begin{aligned}
L_{\mathrm{r}} &= \frac{1}{N_x N_Z}\left\|\mathbf{\Phi}(x)\partial_{tt}\mathbf{A}(t,\boldsymbol{\mu}) - p\partial_{xx}\mathbf{\Phi}(x)\mathbf{A}(t,\boldsymbol{\mu})\right\|_F^2, \\
L_{\mathrm{b}} &= \frac{1}{N_Z}\left\|(\mathbf{\Phi}(0)-\mathbf{\Phi}(1))\mathbf{A}(t,\boldsymbol{\mu})\right\|_F^2 + \frac{1}{N_Z}\left\|(\partial_x\mathbf{\Phi}(0)-\partial_x\mathbf{\Phi}(1))\mathbf{A}(t,\boldsymbol{\mu})\right\|_F^2, \\
L_{\mathrm{i}} &= \frac{1}{N_x N_\mu}\left\|\mathbf{\Phi}(x)\mathbf{A}(0,\boldsymbol{\mu}) - A\sin(2\pi x)\cos(\theta)\right\|_F^2 + \frac{1}{N_x N_\mu}\left\|\mathbf{\Phi}(x)\partial_t\mathbf{A}(0,\boldsymbol{\mu}) + A\omega\sin(2\pi x)\sin(\theta)\right\|_F^2. \\
L_{\mathrm{orth}} &= \max\left(0, \frac{1}{n_{\mathrm{b}}^2}\left\|\mathbf{G}-\mathbf{I}_d\right\|_F^2 - \gamma_{\mathrm{orth}}\right),
\end{aligned}
$$

and $\lambda_{\mathrm{r}}$, $\lambda_{\mathrm{b}}$, $\lambda_{\mathrm{i}}$ and $\lambda_{\mathrm{orth}}$ are 1, 5, 100 and 0.1, respectively. The SpatialNet and TPNet parameters, $\{\theta_x, \theta_z\}$, are then optimized using Adam until convergence, yielding the trained $\mathbf{\Phi}$ and the corresponding approximate solutions.

b) HIDON. We precompute and store the sparse basis evaluations $\mathbf{\Phi}_{\mathrm{s}}$ and their associated spatial differential operator $\mathbf{L\Phi}_{\mathrm{s}} = \{\partial_x\mathbf{\Phi}_{\mathrm{s}}, \partial_{xx}\mathbf{\Phi}_{\mathrm{s}}\}$ on the sparse spatial sampling points $x_{\mathrm{s}} \in \mathcal{X}_{\mathrm{s}} = \{x_i\}_{i=1}^{n_{\mathrm{s}}} \subset \Omega_x$. The weighted loss $L(\theta_z)$ for wave equation is

$$
L(\theta_z) = \lambda_{\mathrm{r}} L_{\mathrm{r}} + \lambda_{\mathrm{b}} L_{\mathrm{b}} + \lambda_{\mathrm{i}} L_{\mathrm{i}}
$$

with

$$
\begin{aligned}
L_{\mathrm{r}} &= \frac{1}{n_{\mathrm{s}} N_Z}\left\|\mathbf{\Phi}_{\mathrm{s}}\frac{d^2\mathbf{A}}{dt^2} - p\partial_{xx}\mathbf{\Phi}_{\mathrm{s}}\mathbf{A}\right\|_F^2, \\
L_{\mathrm{b}} &= \frac{1}{N_Z}\left\|(\mathbf{\Phi}_{\mathrm{s}}(0)-\mathbf{\Phi}_{\mathrm{s}}(1))\mathbf{A}(t,\boldsymbol{\mu})\right\|_F^2 + \frac{1}{N_Z}\left\|(\partial_x\mathbf{\Phi}_{\mathrm{s}}(0)-\partial_x\mathbf{\Phi}_{\mathrm{s}}(1))\mathbf{A}(t,\boldsymbol{\mu})\right\|_F^2, \\
L_{\mathrm{i}} &= \frac{1}{n_{\mathrm{s}} N_\mu}\left\|\mathbf{\Phi}_{\mathrm{s}}\mathbf{A}(0,\boldsymbol{\mu}) - A\sin(2\pi x_{\mathrm{s}})\cos(\theta)\right\|_F^2 + \frac{1}{n_{\mathrm{s}} N_\mu}\left\|\mathbf{\Phi}_{\mathrm{s}}\mathbf{A}(0,\boldsymbol{\mu}) + A\omega\sin(2\pi x_{\mathrm{s}})\sin(\theta)\right\|_F^2,
\end{aligned}
$$

and $\lambda_{\mathrm{r}}$, $\lambda_{\mathrm{b}}$ and $\lambda_{\mathrm{i}}$ are 1, 5 and 100, respectively. After defining $L(\theta_z)$, we update $\theta_z$ using Adam until convergence, obtaining the trained $\mathbf{A}$ and the corresponding approximate solutions.

### 3.2.2 Results and analysis

We again adopt a deliberately sparse solve–enrich–infer configuration (see Fig. 8b and Table 1). The high-resolution spatial and temporal collocation set uses $N_x = 40{,}000$, $N_t = 1{,}000$, with uniform sampling. HIDON and DIDON achieve relative L2 errors of 0.20%±0.08% and 0.18%±0.07%, respectively, while DIDON maintains uniformly low absolute error over representative inference cases (see Fig. 8a–d and Fig. 9). Under comparable memory usage, PIDON converges about one order of magnitude faster in iteration count and achieves roughly three orders of magnitude reduction in wall-clock time compared with PINN (see Fig. 8e–f). HIDON, at a cost slightly lower than PIDON, enriches latent dynamical features over the 77 parameter cases (see Fig. 8f). These trends are consistent with the complexity analysis and mirror those observed in diffusion-reaction equation. The per-case memory, time, and error statistics in Fig. 8g further confirm that PHD-SF retains strong generalization and low computational cost for hyperbolic wave dynamics. This again shows that, once the reusable spatial features and differential information are inherited, HIDON and DIDON can efficiently extend the computation from representative physics-informed solving to large-scale cross-parameter inference.

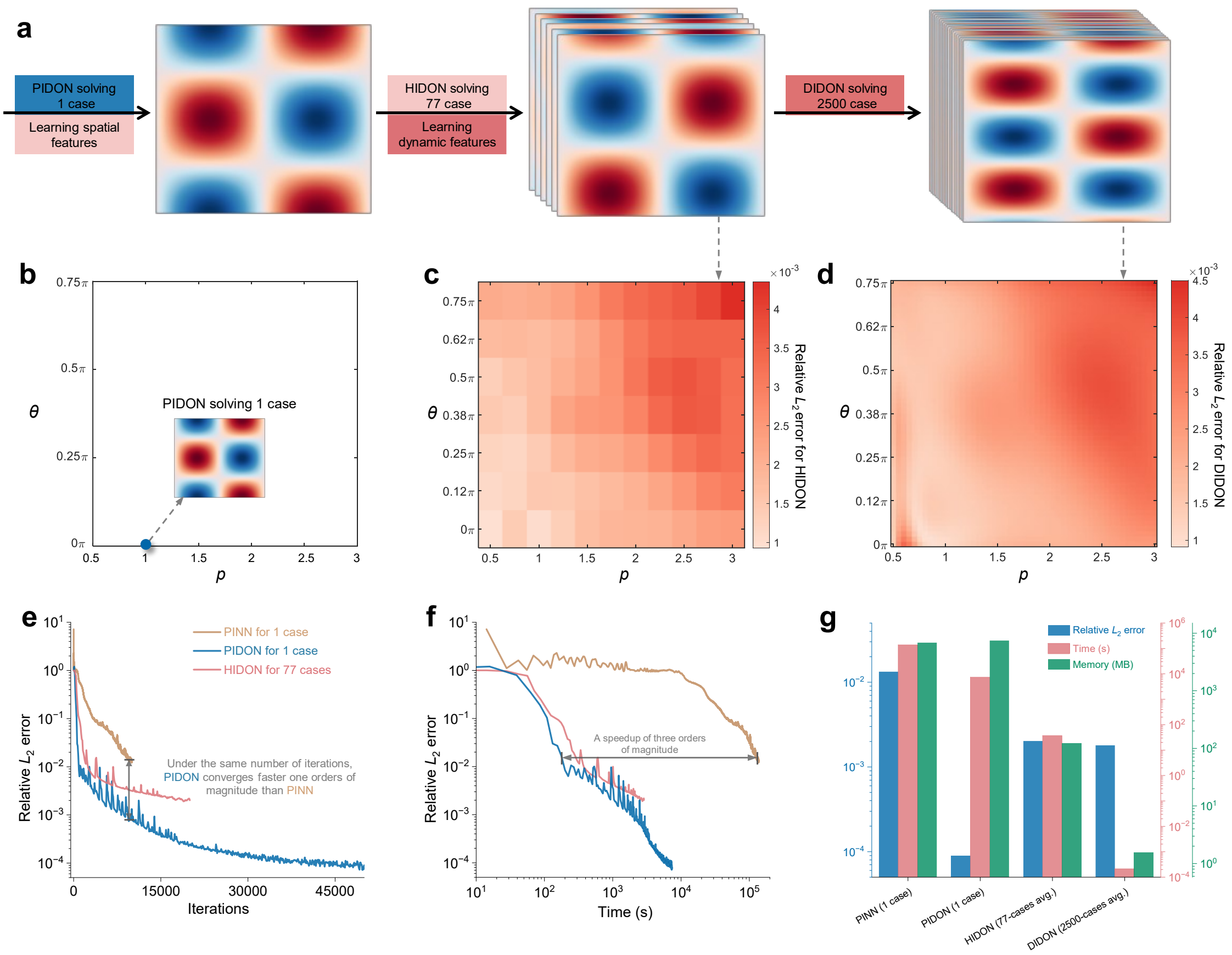

**Fig. 8 | Results for the wave equation. a,** PHD-SF workflow: PIDON is solved on one parameter case while learning spatial features; HIDON then solves 77 parameter cases while learning dynamic features; DIDON finally performs direct inference on 2,500 parameter cases. **b,** PIDON solving case and target parameter domain. **c–d,** Relative $L_2$ error landscapes of HIDON and DIDON over the test domain. **e–f,** Convergence histories of conventional PINN, PIDON and HIDON, reported as relative $L_2$ error versus iteration count and wall-clock time under their respective tasks. **g,** Per-case averages of memory footprint, wall-clock time and relative $L_2$ error for PINN, PIDON, HIDON and DIDON.

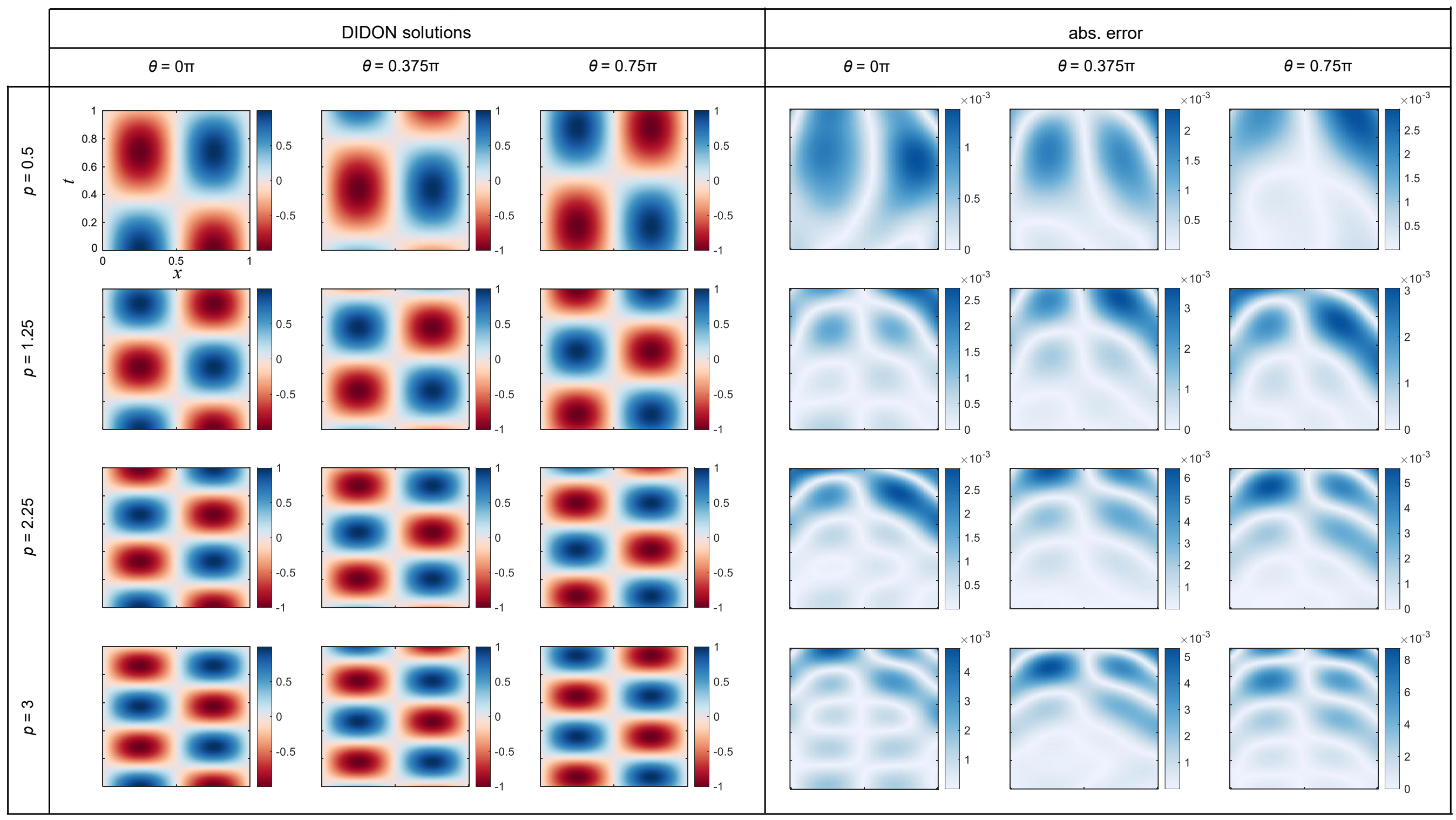

**Fig. 9 | DIDON predicted solution fields and the corresponding absolute error fields for representative parameter cases.**

### 3.3 Burgers' equation

We then consider the viscous Burgers' equation, a canonical nonlinear advection–diffusion model that exhibits steep gradients and shock-like structures,

$$
\begin{aligned}
&\text{PDE: } u_t(x,t) = uu_x(x,t) - vu_{xx}(x,t), \qquad x \in [-1,1], t \in [0,1], \\
&\text{BC: } u(-1,t) = u(1,t) = 0, \quad t \in [0,1], \\
&\text{IC: } u(x,0) = -\sin(\pi x), \qquad x \in [-1,1].
\end{aligned}
$$

here the parameter $v$ controls the viscosity, and thus the balance between convection and diffusion; $v$ itself serves as the system parameter $\boldsymbol{\mu}(v)$.

#### 3.3.1 Solving process

a) PIDON. The weighted loss $L(\theta_x, \theta_z)$ for burger equation reads

$$L(\theta_x, \theta_z) = \lambda_r L_r + \lambda_b L_b + \lambda_i L_i + \lambda_{orth} L_{orth}$$

with

$$
\begin{aligned}
L_{\mathrm{r}} &= \frac{1}{N_x N_Z}\left\|\mathbf{\Phi}(x)\partial_t \mathbf{A}(t,\boldsymbol{\mu}) - (\mathbf{\Phi}(x)\mathbf{A}(t,\boldsymbol{\mu}))\odot(\partial_x \mathbf{\Phi}(x)\mathbf{A}(t,\boldsymbol{\mu})) + v\partial_{xx}\mathbf{\Phi}(x)\mathbf{A}(t,\boldsymbol{\mu})\right\|_F^2,\\
L_{\mathrm{b}} &= \frac{1}{N_Z}\left\|\mathbf{\Phi}(-1)\mathbf{A}(t,\boldsymbol{\mu})\right\|_F^2 + \frac{1}{N_Z}\left\|\mathbf{\Phi}(1)\mathbf{A}(t,\boldsymbol{\mu})\right\|_F^2,\\
L_{\mathrm{i}} &= \frac{1}{N_x N_\mu}\left\|\mathbf{\Phi}(x)\mathbf{A}(0,\boldsymbol{\mu}) + \sin(\pi x)\right\|_F^2.\\
L_{\mathrm{orth}} &= \max\left(0, \frac{1}{n_{\mathrm{b}}^2}\left\|\mathbf{G}-\mathbf{I}_d\right\|_F^2 - \gamma_{\mathrm{orth}}\right),
\end{aligned}
$$

and $\lambda_{\mathrm{r}}$, $\lambda_{\mathrm{b}}$, $\lambda_{\mathrm{i}}$ and $\lambda_{\mathrm{orth}}$ are 1, 5, 100 and 0.1, respectively. The symbol $\odot$ denotes the Hadamard (element-wise) product**.** After defining $L(\theta_x,\theta_z)$, we update $\{\theta_x,\theta_z\}$ using Adam until convergence, obtaining the trained $\mathbf{\Phi}$ and the corresponding approximation $\boldsymbol{u}(\boldsymbol{x},t,\boldsymbol{\mu};\theta_x,\theta_z)$.

b) HIDON. We precompute and store the sparse basis evaluations $\mathbf{\Phi}_{\mathrm{s}}$ and their associated spatial differential operator $\mathbf{L\Phi}_{\mathrm{s}} = \{\partial_x \mathbf{\Phi}_{\mathrm{s}}, \partial_{xx}\mathbf{\Phi}_{\mathrm{s}}\}$ on the sparse spatial sampling points $x_{\mathrm{s}} \in \mathcal{X}_{\mathrm{s}} = \{x_i\}_{i=1}^{n_{\mathrm{s}}} \subset \Omega_x$. The weighted loss $L(\theta_z)$ for burger equation is

$$
L(\theta_z) = \lambda_{\mathrm{r}} L_{\mathrm{r}} + \lambda_{\mathrm{b}} L_{\mathrm{b}} + \lambda_{\mathrm{i}} L_{\mathrm{i}}
$$

with

$$
\begin{aligned}
L_{\mathrm{r}} &= \frac{1}{n_{\mathrm{s}} N_Z}\left\|\mathbf{\Phi}_{\mathrm{s}}\partial_t \mathbf{A} - (\mathbf{\Phi}_{\mathrm{s}}\mathbf{A})\odot(\partial_x \mathbf{\Phi}_{\mathrm{s}}\mathbf{A}) + v\partial_{xx}\mathbf{\Phi}_{\mathrm{s}}\mathbf{A}\right\|_F^2,\\
L_{\mathrm{b}} &= \frac{1}{N_Z}\left\|\mathbf{\Phi}_{\mathrm{s}}(-1)\mathbf{A}(t,\boldsymbol{\mu})\right\|_F^2 + \frac{1}{N_Z}\left\|\mathbf{\Phi}_{\mathrm{s}}(1)\mathbf{A}(t,\boldsymbol{\mu})\right\|_F^2,\\
L_{\mathrm{i}} &= \frac{1}{n_{\mathrm{s}} N_\mu}\left\|\mathbf{\Phi}_{\mathrm{s}}\mathbf{A}(0,\boldsymbol{\mu}) + \sin(\pi x_{\mathrm{s}})\right\|_F^2.
\end{aligned}
$$

and $\lambda_{\mathrm{r}}$, $\lambda_{\mathrm{b}}$ and $\lambda_{\mathrm{i}}$ are 1, 5 and 100, respectively. After defining $L(\theta_z)$, we update $\theta_z$ using Adam until convergence, obtaining the trained $\mathbf{A}$ and the corresponding approximate solutions.

### 3.3.2 Results and analysis

We again adopt a deliberately sparse solve–enrich–infer configuration (see Fig. 10a and Table 1). The high-resolution space–time collocation set uses $N_x = 40{,}000$ spatial, $N_t = 1{,}000$ temporal collocation points under uniform sampling. HIDON and DIDON achieve relative L2 errors of 0.18%±0.26% and 0.14%±0.18%, respectively, and DIDON again shows uniformly low absolute error on representative inference cases (see Fig. 10a–c). Compared with PINN under similar memory usage, PIDON converges slightly faster in iteration count and reduces wall-clock time by roughly two orders of magnitude (Fig. 10d–e). HIDON enriches latent dynamical features over 48 parameter cases at a cost slightly lower than that of the PIDON solving stage (see Fig. 10e). These observations are fully consistent with the theoretical complexity analysis and with the trends observed in the previous examples. The per-case memory, time, and error statistics in Fig. 10f further confirm that PHD-SF retains strong cross-parameter generalization and low computational cost for nonlinear Burgers-type dynamics.

### 3.3.3 Sparse spatial sampling effects for HIDON

To further examine the influence of sparse spatial sampling on HIDON, we conduct an additional study using the Burgers' equation with viscosity v = 0.02. Two sparse spatial sampling strategies are compared: uniform sampling and non-uniform sampling. The non-uniform points are generated from a Gaussian-biased density function, $p(x) = p_0 + \exp[-(x - x_c)^2 / (2\sigma^2)]$, where $x_c$ = 0.25, $\sigma$ = 0.5, and $\rho_0$ = 0.5 in this test. This setting provides a mild non-uniform distribution, which increases the sampling density over a broad dynamically active region while maintaining global coverage of the spatial domain, as shown in Fig. 10i.

The results in Fig. 10g show that the relative $L_2$ error decreases as the sampling ratio $n_s/n_b$ increases. This is expected because a larger number of sparse points provides more governing-equation constraints and improves the identifiability of the latent dynamics. However, after $n_s/n_b$ reaches approximately 10, the error gradually enters a plateau, indicating that the system has become sufficiently constrained and that further increasing ns brings only limited accuracy improvement. The comparison between the two sampling strategies shows that non-uniform sampling is particularly beneficial when $n_s$ is small. As $n_s$ increases, the difference between uniform and non-uniform sampling becomes smaller, because both strategies eventually provide enough governing-equation constraints to stabilize the latent system.

Fig. 10h further shows that the GPU memory requirement increases with $n_s$. This is because more sparse spatial points require more governing-equation evaluations and larger tensors for constructing the reduced latent system. Therefore, the practical selection of $n_s$ should balance accuracy, stability, and computational cost. In general, more complex systems require more sparse spatial points to obtain stable constraints on the reduced latent system. This is particularly evident for the Burgers' equation, where steep-gradient or shock-like solution structures make HIDON more sensitive to both the number and the distribution of sparse sampling points.

It should be noted that the present non-uniform sampling strategy is not claimed to be optimal. It is used here only to demonstrate that the spatial distribution of sparse residual points can affect the information efficiency of HIDON, especially in the small-$n_s$ regime. More systematic sampling strategies, such as adaptive sampling, greedy point selection, or error-indicator-guided sampling, will be investigated in future work.

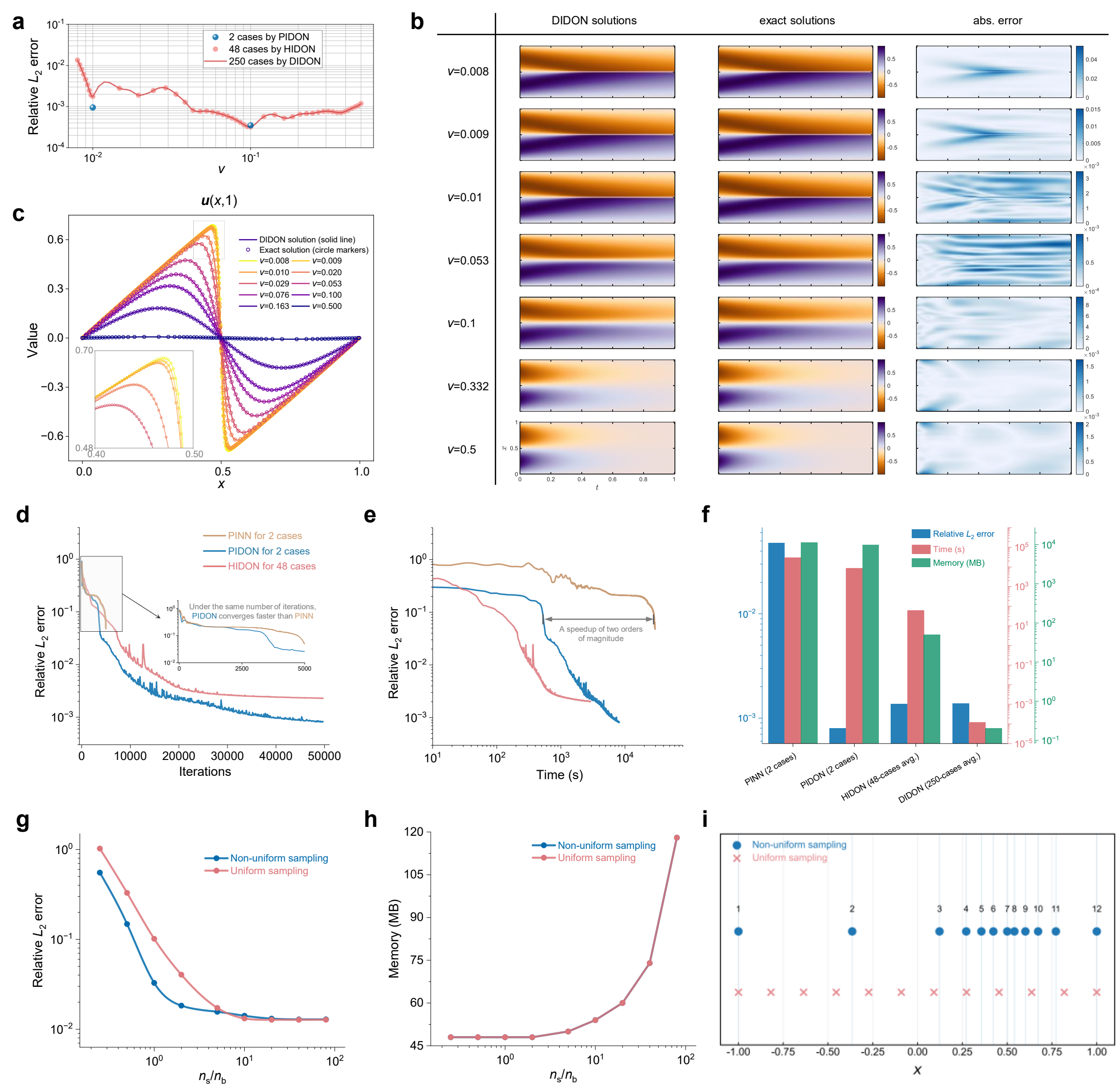

**Fig. 10 | Results for the Burgers' equation. a,** PIDON solving case and target parameter domain, together with the relative $L_2$ errors of PIDON, HIDON and DIDON over the target parameter domain. **b,** For representative parameter cases, DIDON predicted solution fields, the corresponding exact solution fields, and absolute error fields. **c,** Solution fields at $t$ = 1 comparing DIDON predictions with the exact solution for representative parameter cases. **d-e,** Convergence histories of conventional PINN, PIDON and HIDON, reported as relative $L_2$ error versus iteration count and wall-clock time under their respective tasks. **f,** Per-case averages of memory footprint, wall-clock time and relative $L_2$ error for PINN, PIDON, HIDON and DIDON. **g**, Relative $L_2$ error under different sampling ratio $n_s/n_b$ for uniform and non-uniform sparse residual sampling. **h**, GPU memory usage under different sampling ratios. **i**, Representative uniform and non-uniform sparse point distributions for $n_s$ = 12.

## 3.4 Heat transfer equation

We finally consider a two-dimensional transient heat transfer problem, representative of diffusion-dominated thermal processes in solid components. The governing equation reads

$$
\begin{aligned}
&\text{PDE: } u_t = \alpha(u_{xx} + u_{yy}),\\
&\text{BC: } \quad u = t^{\theta}\sin(\pi x), \ (x,y)\in\Gamma_1,\\
&\qquad u = 0, \qquad\qquad (x,y)\in\Gamma_2 \text{ and } \Gamma_4,\\
&\qquad \alpha\nabla u\cdot\mathbf{n} = 0, \quad (x,y)\in\Gamma_3 \text{ and } \Gamma_5,\\
&\text{IC: } \quad u = 0,
\end{aligned}
$$

For the heat transfer equation with boundary partition $\partial\Omega = \Gamma_1 \cup \cdots \cup \Gamma_5$. The detailed computational domain and the boundary partition are shown in Fig. 11(a). Here $\alpha$ controls the thermal diffusivity, while $\theta$ parametrizes a boundary-temperature profile; together they form the system parameter vector $\boldsymbol{\mu} = (\alpha, \theta)$.

### 3.4.1 Solving process

a) PIDON. The weighted loss $L(\theta_x, \theta_z)$ for heat transfer equation reads

$$L(\theta_x,\theta_z) = \lambda_{\mathrm{r}} L_{\mathrm{r}} + \lambda_{\mathrm{b}} L_{\mathrm{b}} + \lambda_{\mathrm{i}} L_{\mathrm{i}} + \lambda_{\mathrm{orth}} L_{\mathrm{orth}}$$

with

$$
\begin{aligned}
L_{\mathrm{r}} &= \frac{1}{N_x N_Z}\left\|\boldsymbol{\Phi}(x,y)\partial_t \mathbf{A}(t,\boldsymbol{\mu}) - \alpha(\partial_{xx}\boldsymbol{\Phi}(x,y) + \partial_{yy}\boldsymbol{\Phi}(x,y))\mathbf{A}(t,\boldsymbol{\mu})\right\|_F^2,\\
L_{\mathrm{b}} &= \frac{1}{N_Z}\left\|\boldsymbol{\Phi}_{\Gamma_1} A(t,\mu)) - g_{\Gamma_1}(t,\theta)\right\|_F^2 + \frac{1}{N_Z}\left\|\boldsymbol{\Phi}_{\Gamma_2} A(t,\mu)\right\|_F^2 + \frac{1}{N_Z}\left\|\boldsymbol{\Phi}_{\Gamma_4} A(t,\mu)\right\|_F^2\\
&\quad + \frac{1}{N_z}\| \alpha\nabla\left(\boldsymbol{\Phi}_{\Gamma_3} A(t,\mu)\right)\cdot\mathbf{n}\|_F^2 + \frac{1}{N_z}\| \alpha\nabla\left(\boldsymbol{\Phi}_{\Gamma_5} A(t,\mu)\right)\cdot\mathbf{n}\|_F^2,\\
L_{\mathrm{i}} &= \frac{1}{N_x N_\mu}\left\|\boldsymbol{\Phi}(x,y)\mathbf{A}(0,\boldsymbol{\mu})\right\|_F^2,\\
L_{\mathrm{orth}} &= \max\left(0, \frac{1}{n_{\mathrm{b}}^2}\left\|\mathbf{G} - \mathbf{I}_d\right\|_F^2 - \gamma_{\mathrm{orth}}\right).
\end{aligned}
$$

Here $g_{\Gamma_1}(t,\theta) = t^{\theta}\sin(\pi x)$ on $\Gamma_1$, and $\lambda_{\mathrm{r}}$, $\lambda_{\mathrm{b}}$, $\lambda_{\mathrm{i}}$ and $\lambda_{\mathrm{orth}}$ are 1, 5, 100 and 0.1, respectively. After defining $L(\theta_x,\theta_z)$, we update $\{\theta_x,\theta_z\}$ using Adam until convergence, obtaining the trained $\boldsymbol{\Phi}$ and the corresponding approximation $\boldsymbol{u}(\boldsymbol{x}, t, \boldsymbol{\mu}; \theta_x, \theta_z)$.

b) HIDON. We precompute and store the sparse basis evaluations $\boldsymbol{\Phi}_{\mathrm{s}}$ and their associated spatial differential operator $\mathbf{L}\boldsymbol{\Phi}_{\mathrm{s}} = \{\partial_x\boldsymbol{\Phi}_{\mathrm{s}}, \partial_y\boldsymbol{\Phi}_{\mathrm{s}}, \partial_{xx}\boldsymbol{\Phi}_{\mathrm{s}}, \partial_{yy}\boldsymbol{\Phi}_{\mathrm{s}}\}$ on the sparse spatial sampling points $x_{\mathrm{s}} \in \mathcal{X}_{\mathrm{s}} = \{x_i\}_{i=1}^{n_{\mathrm{s}}} \subset \Omega_x$. The weighted loss $L(\theta_z)$ for heat transfer equation is

$$L(\theta_z) = \lambda_{\mathrm{r}} L_{\mathrm{r}} + \lambda_{\mathrm{b}} L_{\mathrm{b}} + \lambda_{\mathrm{i}} L_{\mathrm{i}}$$

with

$$
\begin{aligned}
L_{\mathrm{r}} &= \frac{1}{n_{\mathrm{s}}N_Z}\left\|\mathbf{\Phi}_{\mathrm{s}}\partial_t\mathbf{A}(t,\boldsymbol{\mu})-\alpha(\partial_{xx}\mathbf{\Phi}_{\mathrm{s}}+\partial_{yy}\mathbf{\Phi}_{\mathrm{s}})\mathbf{A}(t,\boldsymbol{\mu})\right\|_F^2, \\
L_{\mathrm{b}} &= \frac{1}{N_Z}\left\|\mathbf{\Phi}_{\mathrm{s},\Gamma_1}A(t,\mu))-g_{\mathrm{s},\Gamma_1}(t,\theta)\right\|_F^2+\frac{1}{N_Z}\left\|\mathbf{\Phi}_{\mathrm{s},\Gamma_2}A(t,\mu)\right\|_F^2+\frac{1}{N_Z}\left\|\mathbf{\Phi}_{\mathrm{s},\Gamma_4}A(t,\mu)\right\|_F^2 \\
&\quad+\frac{1}{N_z}\|\alpha\nabla\left(\mathbf{\Phi}_{\mathrm{s},\Gamma_3}A(t,\mu)\right)\cdot\mathbf{n}\|_F^2+\frac{1}{N_z}\|\alpha\nabla\left(\mathbf{\Phi}_{\mathrm{s},\Gamma_5}A(t,\mu)\right)\cdot\mathbf{n}\|_F^2, \\
L_{\mathrm{i}} &= \frac{1}{n_{\mathrm{s}}N_{\mu}}\left\|\mathbf{\Phi}_{\mathrm{s}}\mathbf{A}(0,\boldsymbol{\mu})\right\|_F^2.
\end{aligned}
$$

Here $\mathbf{\Phi}_{\mathrm{s},\Gamma_i}$ denotes the restriction of $\mathbf{\Phi}_{\mathrm{s}}$ to the sparse boundary subset $\mathcal{X}_{\mathrm{s}}\cap\Gamma_i$, and $\lambda_{\mathrm{r}}$, $\lambda_{\mathrm{b}}$ and $\lambda_{\mathrm{i}}$ are 1, 5 and 100, respectively. After defining $L(\theta_z)$, we update $\theta_z$ using Adam until convergence, obtaining the trained $\mathbf{A}$ and the corresponding approximated solutions.

c) DIDON. DIDON directly inherits the trained SpatialNet from PIDON and the enriched TPNet from HIDON, and performs direct solution inference.

### 3.4.2 Results and analysis

The sparse solve–enrich–infer configuration is summarized in Table 1. The high-resolution space–time collocation set uses $N_{\mathrm{c}}=14,400$ spatial, $N_t=1{,}000$ temporal collocation points under uniform sampling. DIDON predictions agree closely with the reference solutions for representative inference cases (see Fig. 11b–d). HIDON and DIDON achieve relative L2 errors of 0.60%±0.32% and 0.62%±0.41%, respectively. Under comparable memory usage, PIDON converges slightly faster than PINN in iteration count and reduces wall-clock time by roughly two orders of magnitude, while HIDON enriches latent dynamical features over 108 parameter cases at a cost slightly below that of PIDON (see Fig. 11e–f). These observations are fully consistent with the theoretical complexity analysis and with the trends observed in the previous examples. The per-case statistics in Fig. 11g further confirm the strong generalization and low computational cost of PHD-SF for two-dimensional heat transfer problems.

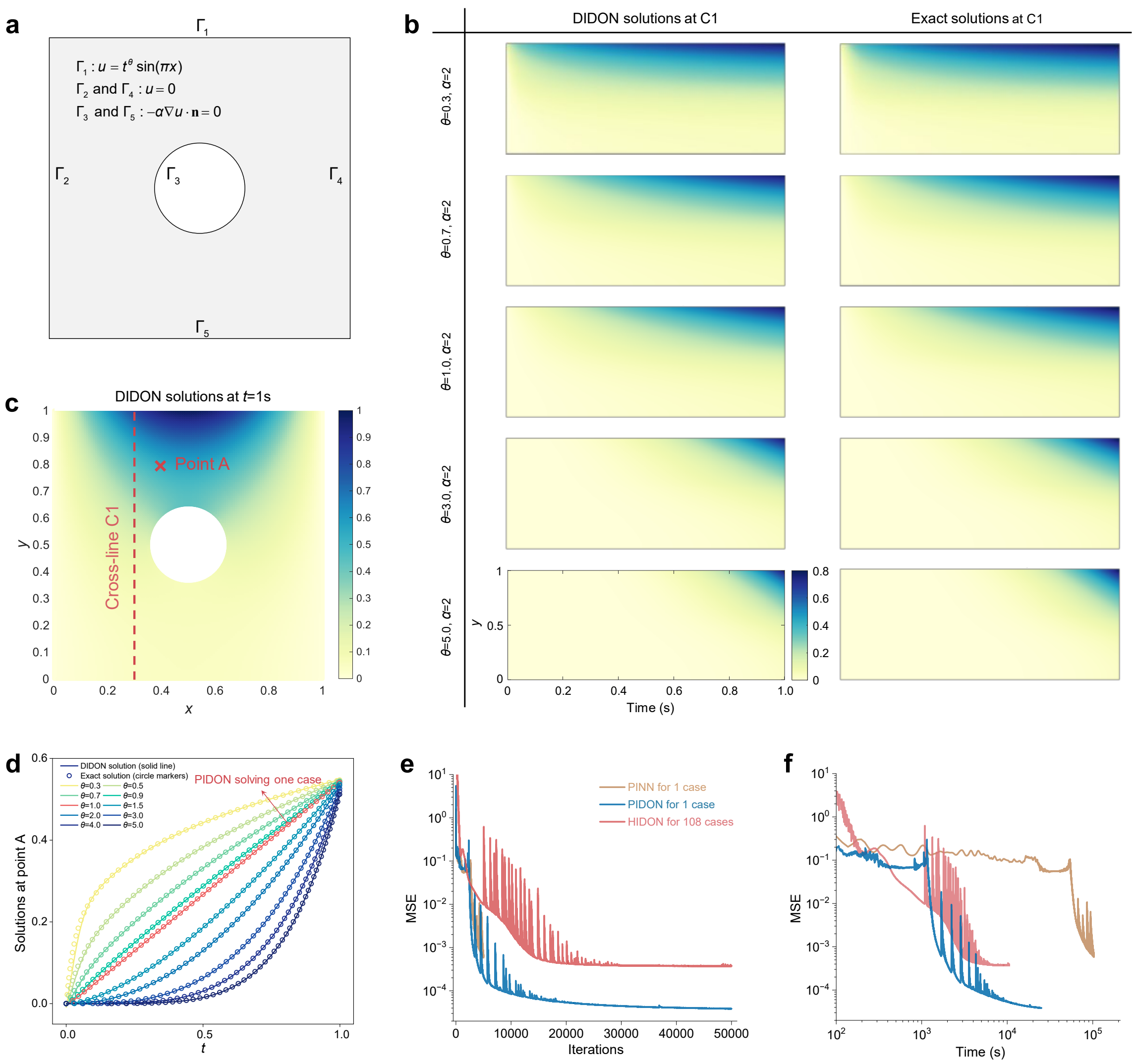

**Fig. 11 | Results for the heat-transfer equation. a,** Computational domain and boundary-condition specification. **b,** DIDON solution and exact solution fields for representative test domain parameter cases, shown on cross-section **C1**. **c,** PIDON solution at $t$ = 1. **d,** DIDON solutions and exact solutions for representative cases in target parameter domain, reported at monitoring point **A**. **e–f,** Convergence histories of conventional PINN, PIDON and HIDON, reported as relative $L_2$ error versus iteration count and wall-clock time under their respective tasks.

## 4. Discussion

This work proposes PHD-SF, a compact physics-to-data spectrum framework for accelerating parametric time-dependent PDEs. Its core contribution lies not simply in constructing three models, namely PIDON, HIDON, and DIDON, but in organizing them as three operating modes of a unified DON-based architecture at different stages of the physics-to-data spectrum, with cross-stage connections established through reusable model information. Unlike the conventional label-mediated workflow of “batch high-fidelity solving–solution-label training–direct inference,” PHD-SF forms a compact solve–enrich–infer acceleration path. This path significantly reduces the computational burden associated with batch high-fidelity solving and labeled-data construction in the conventional label-mediated workflow, providing a new computational mode for fast parametric PDE solving and cross-parameter inference.

The reusable model information in PHD-SF is mainly organized at two levels. First, at the representational-information level, PIDON learns spatial features and the associated spatial differential-operator information during physics-informed solving over a small number of representative parameter cases; HIDON then inherits this information and further enriches the latent dynamical features encoded in the BCFs over a broader time–parameter domain. The resulting spatial features, spatial differential-operator information, and latent dynamical features constitute the main reusable model information in PHD-SF. Second, at the structural-information level, PIDON, HIDON, and DIDON share a unified DON-based architecture and realize cross-stage structural reuse through parameter reuse and staged training and freezing of subnetworks. It is worth emphasizing that the reuse of both representational and structural information is the core mechanism by which PHD-SF achieves compact acceleration across the physics-to-data spectrum.

Moreover, from the broader perspective of PDE solvers, PHD-SF can be understood as a label-free physics-informed solver with cross-parameter generalization capability. It requires neither precomputed high-fidelity solution snapshots as training labels nor explicit solution-field label transfer among PIDON, HIDON, and DIDON. Similar to traditional solvers such as FEM and PINN, PHD-SF can perform high-accuracy physics-constrained solving according to the given governing equations. The key difference is that, after solving a single case or a small number of cases, traditional solvers can hardly acquire direct cross-parameter generalization capability. By contrast, PHD-SF can subsequently perform accurate and fast solving over a nearby or enriched parameter domain using HIDON and DIDON, thereby providing cross-parameter generalization capability. This ability to extend from single-case solving to cross-parameter generalized solving is a key capability that current deep-learning-based PDE solvers urgently need to achieve.

The results on four benchmark time-dependent PDE problems demonstrate that PHD-SF enables effective cross-parameter solving under sparse, one-shot-like parameter configurations. The numerical results show that the framework maintains high accuracy in physics-constrained solving for representative cases and achieves reliable generalized inference for unseen parameter cases. Meanwhile, the total end-to-end computational cost of PHD-SF is lower than that required to train a conventional PINN for a single parameter case, while providing cross-parameter direct-inference capability that is typically unavailable in single-case PINN solvers. These results indicate that PHD-SF not only realizes spectrum-wide modeling from physics-constrained to data-informed modes, but also provides an efficient and generalizable deep-learning-based solution path for many-query parametric PDE problems.

Beyond its methodological significance, PHD-SF is broadly applicable to time-dependent parametric PDE systems arising in fluid mechanics, heat transfer, solid mechanics, and related fields. The accuracy–cost–generalization performance illustrated in Fig. 5i supports a wide range of engineering demands, including design optimization, online monitoring, and digital twins. In particular, digital-twin applications typically require fit-for-purpose integration and trade-offs among physics-based, data-driven, and hybrid models; therefore, PHD-SF offers a potential model foundation for such applications [41]. Moreover, PHD-SF is compatible with existing learning paradigms: operator-learning modules may strengthen DIDON, advanced PINN variants may improve the robustness of PIDON, and transfer learning or knowledge distillation may further enhance adaptability and deployment efficiency.

Although the present study demonstrates the effectiveness of PHD-SF on four representative time-dependent PDE benchmarks, several extensions remain open. First, the sparse spatial points in HIDON are selected using a simple uniform baseline. This choice is sufficient to demonstrate the hyper-reduction-like effect of sparse governing-equation enforcement, but more adaptive sampling strategies may further improve robustness and efficiency, especially for strongly localized or shock-like dynamics. Second, the current DIDON inference is performed within the temporal–parametric domain enriched by HIDON. Extending it to out-of-domain extrapolation requires additional transfer or adaptive enrichment mechanisms. Third, the present work focuses on regular benchmark problems; extending PHD-SF to complex geometries, unstructured discretizations, and coupled multi-physics systems is an important direction for future work. Finally, although the closure of HIDON is discussed through dimension counting and local identifiability of the sparse constraint Jacobian, rigorous stability analysis and a priori/a posteriori error estimates remain to be developed.

# Appendix

## S1 Unified-mode interpretation of PHD-SF

Although the main text describes PHD-SF as comprising PIDON, HIDON, and DIDON for convenience of presentation, these should more fundamentally be understood as three operating modes of a single DON-based framework. The network architecture itself remains unchanged throughout the workflow; what changes across stages is only the trainable part of the network.

In the PIDON mode, both SpatialNet and TPNet are optimized to solve a small set of representative high-fidelity PDE cases while learning transferable spatial features and their associated differential information. After PIDON training, SpatialNet is frozen, and the framework switches to the HIDON mode, in which only TPNet is further updated to solve reduced ODE systems and enrich latent dynamical features over a larger set of parameter cases. After HIDON training, TPNet is further frozen, yielding the DIDON mode for direct cross-parameter inference without any further parameter update.

Under this view, the inheritance mechanism in PHD-SF is implemented as staged parameter freezing and mode switching within a unified architecture, rather than as information transfer among three entirely separate models. The essential distinction among PIDON, HIDON, and DIDON therefore lies not in different network structures, but in different trainable subnetworks, levels of governing-equation enforcement, and computational roles across the physics-to-data spectrum.

## S2 Proof of Proposition1

***Proposition1****(Density of the PIDON separable function class)***.** Let $\Omega_x \subset \mathbb{R}^{d_x}$ and $\Omega_z \subset \mathbb{R}^{d_z}$ be compact sets, and define the PIDON-associated separable function family

$$\mathcal{F} := \bigcup_{n\geq 1}\mathcal{F}_n, \quad \mathcal{F}_n := \left\{ \hat{\boldsymbol{u}}(\boldsymbol{x},\boldsymbol{z}) = \sum_{i=1}^{n} \varphi_i(\boldsymbol{x}) a_i(\boldsymbol{z}) \mid \varphi_i \in C(\Omega_x), a_i \in C(\Omega_z) \right\} \subset C(\Omega_x \times \Omega_z),$$

equipped with the uniform norm $\|\cdot\|_\infty$. Then $\bigcup_{n\geq 1}\mathcal{F}_n$ is dense in $C(\Omega_x \times \Omega_z)$. In other words, PIDON possesses a uniform universal approximation property: any continuous parametric PDE solution can be approximated to arbitrary accuracy by a PIDON representation.

***Proof.*** Consider the algebra of finite sums of separable continuous functions on the compact product domain $\Omega_x \times \Omega_z$, $\mathcal{A} := \left\{ \sum_{i=1}^{n} f_i(x) g_i(z) \| \; n \in \mathbb{N}, f_i \in C(\Omega_x), g_i \in C(\Omega_z) \right\}$. It is straightforward that $\mathcal{A}$ is a subalgebra of $C(\Omega_x \times \Omega_z)$ that (i) contains the constant functions, and (ii) separates points on $\Omega_x \times \Omega_z$ (if two points differ in either $x$ or $z$, one may choose a continuous function depending only on that coordinate to distinguish them). By the Stone–Weierstrass theorem, $\mathcal{A}$ is dense in $C(\Omega_x \times \Omega_z)$ in the uniform norm [32, 33]. Hence, for any $u \in C(\Omega_x \times \Omega_z)$ and any $\varepsilon > 0$, there exists $\hat{u} \in \mathcal{F}_n$ for some $n$ such that $\| u - \hat{u} \|_\infty < \varepsilon$; that is, finite linear combinations of separable continuous functions uniformly approximate any continuous target function to arbitrary accuracy.

On the other hand, the classical universal approximation theorems for feed-forward neural networks on compact sets state that, with a non-polynomial activation function, a multilayer perceptron can uniformly approximate any continuous function on a compact domain to arbitrary accuracy[34-36]. Therefore, each function $f_i \in C(\Omega_x)$ and $g_i \in C(\Omega_z)$ can be approximated by neural networks $\varphi_i(x)$ and $a_i(z)$, respectively, with arbitrarily small errors. Combining these factorwise approximations with a standard product error bound yields that the induced PIDON approximation can be made uniformly close to $u$ within any prescribed tolerance. Consequently, PIDON possesses a uniform universal approximation property for any continuous parametric PDE solution[16, 17].

### S3 Closure and sparse-point selection for HIDON

From the perspective of degrees of freedom (DoFs) versus constraints, HIDON can be interpreted as solving a dynamical system with $n_\mathrm{b}$ DoFs, subject to $n_\mathrm{s}$ governing equation constraints enforced on a sparse spatial sampling set $\boldsymbol{x}_\mathrm{s} \in \mathcal{X}_\mathrm{s} = \{x_i\}_{i=1}^{n_x} \subset \Omega_x$. Specifically, the governing equations evaluated at the $n_\mathrm{s}$ sampled points act as *system constraints*, while the latent dynamics (parameterised by TPNet) constitute the unknown DoFs to be inferred. This viewpoint naturally raises two fundamental questions:

***Question 1.*** How should the number of sparse sampling points $n_\mathrm{s}$ be chosen to ensure system closure?

***Discussion.*** From the viewpoint of algebraic closure, the sparse constraint system should provide no fewer effective independent constraints on the unknown BFCs $\mathbf{A}(t,\boldsymbol{\mu}) \in \mathbb{R}^n$ than the number of system DoFs $n_\mathrm{b}$. From the perspective of deep-learning-based solving, *system closure* means that, for a given $(t,\boldsymbol{\mu})$, the physical constraints enforced at the sparse control points can uniquely (or at least locally uniquely) determine $\mathbf{A}(t,\boldsymbol{\mu})$, so that the associated optimisation problem admits an identifiable minimiser and exhibits stability with respect to perturbations.

To formalise this idea, we define the sparse constraint vector (under $(t,\boldsymbol{\mu})$) as

$$\mathbf{F}_{t,\mu}(\mathbf{A}) = \left[ \mathcal{C}(x_1,t,\mu;\mathbf{A}), \quad \ldots, \quad \mathcal{C}(x_{n_\mathrm{s}},t,\mu;\mathbf{A}) \right]^\top \in \mathbb{R}^{n_\mathrm{s}},$$

where $\mathcal{C}$ denotes the physics constraint evaluated at a sampling point (e.g., governing-equation residuals and/or boundary/interface constraints, depending on the formulation). A natural local closure condition is that, in a neighbourhood of the target solution $\mathbf{A}^\star$, the constraint mapping is sufficiently sensitive to $\mathbf{A}$. Concretely, the Jacobian $\boldsymbol{J}_{t,\mu}(\mathbf{A}^\star) = (\partial \mathbf{F}_{t,\mu} / \partial \mathbf{A})\mathbf{A}^\star \in \mathbb{R}^{n_\mathrm{s} \times n_\mathrm{b}}$ should have full column rank. Equivalently, the information matrix $\boldsymbol{J}^T\boldsymbol{J}$ should be well-conditioned in the sense that its smallest eigenvalue is strictly positive: $\lambda_{\min}(\boldsymbol{J}^\top\boldsymbol{J}) > 0$. This condition implies that the sparse constraints can locally distinguish different $\mathbf{A}$, thereby preventing an underdetermined system. By a basic dimension-counting argument, a necessary condition for closure is $n_\mathrm{s} \geq n_\mathrm{b}$. However, due to the variation of $(t,\boldsymbol{\mu})$, the nonlinearity of the constraints, and training noise, such ideal identifiability is difficult to guarantee a priori over the entire domain. Therefore, in practice one typically adopts an over-constraint (over-sampling) strategy: $n_\mathrm{s} = \gamma_\mathrm{s} n_\mathrm{b}$, $\gamma_\mathrm{s} > 1$, using redundant constraints to improve robustness and training stability, and to avoid underdetermined or ill-conditioned behaviour when $n_\mathrm{s}$ is too small. In this work, to balance closure with online cost, we choose a moderate $\gamma_\mathrm{s}$ in the numerical examples (typically $\gamma_\mathrm{s} \in [2,20]$), so as to preserve sparsity and the resulting acceleration benefits. Notably, this closure issue—and the corresponding over-sampling remedy—is also standard in numerical reduced-order modelling. Representative examples include over-sampled regression in gappy POD [42], sparse interpolation point selection in DEIM[10], and sparse-constraint DHROM formulations[43]. These mature ROM/HROM practices can be directly borrowed and transferred to HIDON.

***Question 2.*** Given a prescribed number of sampling points $n_{\mathrm{s}}$, how should the sparse sampling set $\mathcal{X}_{\mathrm{s}}$ be selected?

***Discussion.*** With $n_{\mathrm{s}}$ fixed, the optimal sampling pattern can be viewed as one that makes the sparse constraints maximally informative for distinguishing $\mathbf{A}(t,\mu)$ across as broad a subset of the parameter domain as possible. A practical and widely used criterion is to improve the spectral properties of the information matrix $G_z(x_{\mathrm{s}}) = \boldsymbol{J}_z(\mathbf{A})^T \boldsymbol{J}_z(\mathbf{A}) \in \mathbb{R}^{n_{\mathrm{b}} \times n_{\mathrm{b}}}$ over the training cases—e.g., maximising its smallest eigenvalue or reducing its condition number—thereby enhancing identifiability and numerical stability. Since $\boldsymbol{J}$ for unseen parameter instances cannot be known a priori, this objective cannot be strictly enforced beforehand. This challenge is also ubiquitous in hyper-reduction for ROMs, where mature point-selection strategies exist, such as DEIM's greedy selection (and its variants) that target stable sparse interpolation.

Although the determination of $n_{\mathrm{s}}$ and $\mathcal{X}_{\mathrm{s}}$ are both important and interesting, they are not the focus of this work, and a full exploration is beyond the scope of this manuscript. In our numerical experiments, we adopt a simple and reproducible baseline: $\gamma_{\mathrm{s}} \in [2,10]$ and uniform sampling for $\mathcal{X}_{\mathrm{s}}$.

**S4 Construction and optimization of the weighted MSE loss for PIDON**

Based on the separable SD–TPD representation and the differentiation-transfer strategy described in the main text, we construct a weighted mean-squared-error (MSE) objective on the spatial (SD) and temporal–parametric (TPD) collocation sets using $\mathbf{\Phi}$, $\mathbf{L\Phi}$, $\mathbf{A}$, and $\partial_t \mathbf{A}$. Specifically, we define：

$$L(\theta_x, \theta_z) = \lambda_{\mathrm{r}} L_{\mathrm{r}} + \lambda_{\mathrm{b}} L_{\mathrm{b}} + \lambda_{\mathrm{i}} L_{\mathrm{i}} + \lambda_{\mathrm{orth}} L_{\mathrm{orth}},$$

where $\theta_x$ and $\theta_z$ denote the parameters of SpatialNet and TPNet, respectively, and $\lambda_{\mathrm{r}}$, $\lambda_{\mathrm{b}}$, $\lambda_{\mathrm{i}}$, $\lambda_{\mathrm{orth}}$ are the corresponding weights.

a) Interior PDE residual. On the interior collocation points $\boldsymbol{x} \in \Omega_x$, the interior PDE residual is constructed by

$$L_{\mathrm{r}} = \mathrm{MSE}(\boldsymbol{\Phi}\partial_t \mathbf{A} - \mathcal{N}(\boldsymbol{\Phi}\mathbf{A}, (\mathbf{L}\boldsymbol{\Phi})\mathbf{A}); \boldsymbol{x}, t, \boldsymbol{\mu})), \quad \boldsymbol{x} \in \Omega_x,$$

where $\mathcal{N}$ denotes the (possibly nonlinear) PDE operator written in residual form.

b) Boundary condition residual. On boundary collocation points $\boldsymbol{x}_{\mathrm{BC}}$ on $\partial\Omega_x$, we enforce the boundary operator $\mathcal{B}(\cdot)$ by

$$L_{\mathrm{b}} = \mathrm{MSE}(\mathcal{B}\left(\boldsymbol{\Phi}(\boldsymbol{x}_{\mathrm{BC}})\mathbf{A}(t, \boldsymbol{\mu})\right) - g(\boldsymbol{x}_{\mathrm{BC}}, t, \boldsymbol{\mu})),$$

where $g$ is the prescribed boundary condition.

c) Initial condition residual. On spatial points $\boldsymbol{x}$ on $\Omega_x$ at $t$=0, we impose

$$L_{\mathrm{i}} = \mathrm{MSE}(\mathcal{B}\left(\boldsymbol{\Phi}(\boldsymbol{x})\mathbf{A}(0, \boldsymbol{\mu})\right) - u_0(\boldsymbol{x}, \boldsymbol{\mu})),$$

where $u_0$ denotes the initial condition. Here, $L_{\mathrm{r}}$, $L_{\mathrm{b}}$, and $L_{\mathrm{i}}$ correspond to the interior PDE, boundary conditions, and initial conditions, respectively. Their weights $\lambda_{\mathrm{r}}$, $\lambda_{\mathrm{b}}$, $\lambda_{\mathrm{i}}$, balance the relative importance of these constraints during training.

d) Orthogonality regularization. To improve the quality of the learned $\boldsymbol{\Phi}$ while maintaining numerical stability, we introduce an orthogonality regulariser

$$L_{\mathrm{orth}} = \mathrm{MSE}(\boldsymbol{G} - \boldsymbol{I}_d),\ \boldsymbol{G}_{ij} = \left\langle \varphi_i, \varphi_j \right\rangle = \sum_{k=1}^{N} \left\langle \varphi_i(k), \varphi_j(k) \right\rangle,$$

where $\boldsymbol{G}$ is the (discrete) Gram matrix computed on $\Omega_x$, and $\boldsymbol{I}_d$ is the identity matrix. Intuitively, encouraging $\boldsymbol{G} \approx \boldsymbol{I}_d$ promotes approximate orthogonality among $\boldsymbol{\Phi}$, which enlarges the effective subspace and can improve the expressive capacity and cross-parameter generalisation of downstream HIDON and DIDON. This design echoes classical POD, where orthogonal bases are sought to improve subspace quality. Since enforcing strict orthogonality may increase optimisation difficulty and computational burden, we set $\lambda_{\mathrm{orth}}$ to be typically one order of magnitude smaller than the PDE weight $\lambda_{\mathrm{r}}$, alleviating early-stage convergence pressure. Moreover, because $\lambda_{\mathrm{orth}}$ is often difficult to reduce to (near) zero in practice, we introduce an orthogonality tolerance $\gamma_{\text{orth}}$ and rewrite the regulariser as

$$L_{\text{orth}} = \max\left(0, \text{MSE}(\boldsymbol{G} - \boldsymbol{I}_d) - \gamma_{\text{orth}}\right).$$

Accordingly, when $\text{MSE}(\boldsymbol{G} - \boldsymbol{I}_d) \le \gamma_{\text{orth}}$, the $\boldsymbol{\Phi}$ is considered “sufficiently orthogonal”. The tolerance $\gamma_{\text{orth}}$ can also be made dependent on the number of basis functions to reflect the fact that strict orthogonality becomes harder to enforce as the basis dimension increases. Overall, $L_{\text{orth}}$ provides a flexible trade-off: stronger orthogonality constraints may raise the initial training cost, but often yield a richer and more representative subspace, benefiting subsequent modelling and generalization.

e) Optimization. After defining $L(\theta_x, \theta_z)$, we compute gradients via backpropagation through both SpatialNet and TPNet, and update $\{\theta_x, \theta_z\}$ using Adam (or other optimizers) until convergence. This yields the trained networks and the corresponding PIDON approximation solution $u$.

**S5. Construction and optimization of the weighted MSE loss for HIDON**

Following the sparse spatial sampling idea from HROMs, we restrict the enforcement of the governing equations to a sparse spatial sampling set $\boldsymbol{x}_{\text{s}} \in \mathcal{X}_{\text{s}} = \{x_i\}_{i=1}^{n_x} \subset \Omega_x$. Accordingly, we precompute and store the sparse basis evaluations $\boldsymbol{\Phi}_{\text{s}}$ and their associated spatial differential operator $\mathbf{L}\boldsymbol{\Phi}_{\text{s}}$, and construct a weighted MSE loss on the sparse spatial set $\mathcal{X}_{\text{s}}$ and the temporal–parametric collocation set $\mathcal{Z}_{\text{h}}$. Concretely,

$$L(\theta_z) = \lambda_{\text{r}} L_{\text{r}} + \lambda_{\text{b}} L_{\text{b}} + \lambda_{\text{i}} L_i,$$

where $L_{\text{r}}$, $L_{\text{b}}$, and $L_{\text{i}}$ denote the residual losses for the reduced dynamics, boundary conditions, and initial conditions, respectively, and $\lambda_{\text{r}}$, $\lambda_{\text{b}}$ and $\lambda_{\text{i}}$, are their weights.

a) Interior sparse equation residual. On the interior sparse collocation points $\boldsymbol{x}_{\text{s}} \in \mathcal{X}_{\text{s}}$, the interior PDE residual is constructed by

$$L_{\text{r}} = \text{MSE}\left(\frac{d\mathbf{A}}{dt} - \mathcal{N}(\boldsymbol{\Phi}_{\text{s}}\mathbf{A}, (\mathbf{L}\boldsymbol{\Phi}_{\text{s}})\mathbf{A}; \boldsymbol{x}_{\text{s}}, t, \boldsymbol{\mu})\right), \quad \boldsymbol{x}_{\text{s}} \in \mathcal{X}_{\text{s}}$$

where $\mathcal{N}$ denotes the governing operator.

b) Boundary-condition residual. On boundary collocation points $\boldsymbol{x}_{\mathrm{BC}} \in \partial\Omega_x \cap \mathcal{X}_{\mathrm{s}}$, we enforce the boundary operator $\mathcal{B}(\cdot)$ by

$$L_{\mathrm{b}} = \mathrm{MSE}(\mathcal{B}\left(\boldsymbol{\Phi}_{\mathrm{s}}(\boldsymbol{x}_{\mathrm{BC}})\mathbf{A}(t,\boldsymbol{\mu})\right) - g(\boldsymbol{x}_{\mathrm{BC}},t,\boldsymbol{\mu})), \quad \boldsymbol{x}_{\mathrm{BC}} \in \partial\Omega_x \cap \mathcal{X}_{\mathrm{s}},$$

where $g$ is the boundary condition.

c) Initial-condition residual. On spatial points $\boldsymbol{x}_{\mathrm{s}} \in \mathcal{X}_{\mathrm{s}}$ at $t$=0, we impose

$$L_i = \mathrm{MSE}(\boldsymbol{\Phi}_{\mathrm{s}}(\boldsymbol{x}_{\mathrm{s}})\mathbf{A}(0,\boldsymbol{\mu}) - u_0(\boldsymbol{x}_{\mathrm{s}},\boldsymbol{\mu})) \quad \boldsymbol{x}_{\mathrm{s}} \in \mathcal{X}_{\mathrm{s}}.$$

where $u_0$ denotes the initial condition.

d) Optimization. In line with standard deep-learning solvers, the evaluated loss $L(\theta_z)$ is passed to an optimizer, and gradient-based methods (e.g., Adam) are used to iteratively update only the parameters of TPNet, $\theta_z$, while SpatialNet is kept frozen throughout HIDON training. Minimizing $L(\theta_z)$ yields a converged parameter set $\theta_z^*$ and the corresponding BCFs $\mathbf{A}(\theta_z^*)$. Finally, inserting the converged BCFs back into the separable representation (Eq. (2) in the main text) produces the final approximate solution.

## S6 Network architecture and Hyper-parameter settings

Network architecture settings are shown in Table 2. And Hyper-parameter settings are shown in Table 3. In all examples considered in this work, the branch net and the trunk net are equipped with hyperbolic tangent activation functions (Tanh). In this work, we tuned these hyper-parameters manually, without attempting to find the absolute best hyper-parameter setting. This process can be automated in the future leveraging effective techniques for meta-learning and hyper-parameter optimization.

**Table 3. Default hyper-parameter settings for each benchmark employed in this work (unless otherwise stated).**

| Governing law | Model | Learning rate | Iterations | Optimizer |
|---|---|---|---|---|
| **Diffusion reaction** | PINN | 0.001 | 10,000 | Adam |
| | PIDON | 0.001 | 50,000 | Adam |

| | HIDON | 0.005 | 10,000 | Adam |
|---|---|---|---|---|
| **Wave** | PINN | 0.01 | 10,000 | Adam |
| | PIDON | 0.01 | 50,000 | Adam |
| | HIDON | 0.01 | 20,000 | Adam |
| **Burger** | PINN | 0.001 | 5,000 | Adam |
| | PIDON | 0.001 | 50,000 | Adam |
| | HIDON | 0.01 | 50,000 | Adam |
| **Heat transfer** | PINN | 0.002 | 5,000 | Adam |
| | PIDON | 0.002 | 50,000 | Adam |
| | HIDON | 0.005 | 50,000 | Adam |

## S7 Explanation of the collocation-set sizes

Table 4 summarizes the sampling sizes and effective network-evaluation scales used in all benchmarks, unless otherwise stated. Here, $N_x$ denotes the number of spatial sampling points in the SD domain, $N_t$ denotes the number of temporal sampling points, and $N_\mu$ denotes the number of sampled parameter cases. We define $N_z = N_t N_\mu$ as the number of temporal–parametric sampling points. For HIDON, $n_s$ denotes the number of sparse spatial control points used for sparse residual enforcement.

The last column of Table 4 reports $N_{eval}$, the effective number of network evaluations associated with governing-equation enforcement. It should not be interpreted as the number of physical collocation pairs in the full tensor-product domain. In a conventional PINN, the network takes the coupled input ($\boldsymbol{x}$,$\boldsymbol{z}$) and must be evaluated and differentiated at every space–time–parameter pair, leading to $N_x N_z$ evaluations. In PIDON, however, SpatialNet and TPNet are evaluated separately on the SD and TPD sampling sets, and the residual is assembled from these separated outputs. Therefore, the effective network-evaluation scale becomes $N_x + N_z$.

Accordingly, $N_{eval} = N_x N_z$ for a conventional PINN, $N_{eval} = N_x + N_z$ for PIDON, $N_{eval} = n_s + N_z$ for HIDON, $N_{eval} = 0$ for DIDON (see Fig. 4 for a schematic illustration). Unless otherwise specified, all spatial points, temporal points, and parameter cases are generated by uniform sampling over their corresponding domains; for HIDON, the sparse set $\mathcal{X}_s$ is also constructed by uniform sampling.

**Table 4. The collocation-set sizes for each benchmark employed in this work (unless otherwise stated).**

| Governing law | Model | $N_x$ | $N_t$ | $N_\mu$ | $n_s$ | $N_{eval}$ |
|---|---|---|---|---|---|---|
| **Diffusion reaction** | PINN | 40,000 | 1,000 | 1 | | $4\times10^7$ |
| | PIDON | 40,000 | 1,000 | 1 | | 41,000 |
| | HIDON | 40,000 | 1,000 | 121 | 20 | 121,020 |
| | DIDON | 40,000 | 1,000 | 2,500 | | 0 |
| **Wave** | PINN | 40,000 | 1,000 | 1 | | $4\times10^7$ |
| | PIDON | 40,000 | 1,000 | 1 | | 41,000 |
| | HIDON | 40,000 | 1,000 | 77 | 20 | 77,020 |
| | DIDON | 40,000 | 1,000 | 2,500 | | 0 |
| **Burger** | PINN | 40,000 | 1,000 | 2 | | $4×10^7$ |
| | PIDON | 40,000 | 1,000 | 2 | | 41,000 |
| | HIDON | 40,000 | 1,000 | 48 | 100 | 48,020 |
| | DIDON | 40,000 | 1,000 | 250 | | 0 |
| **Heat transfer** | PINN | 14,000 | 1,000 | 1 | | $1.4×10^7$ |
| | PIDON | 14,000 | 1,000 | 1 | | 15,000 |
| | HIDON | 14,000 | 1,000 | 108 | 20 | 108,020 |
| | DIDON | 14,000 | 1,000 | 1,500 | | 0 |

## S8 Full-batch training protocol and gradient accumulation for PINN

To ensure a fair comparison across all models and all parameter cases, we adopt a full-batch training protocol throughout: each optimization step is defined with respect to the full set of training collocation points for the corresponding loss terms. For PINN, however, the number of collocation points can be extremely large, making a literal full-batch forward/backward pass infeasible under GPU memory constraints. We therefore implement PINN using gradient accumulation: the full collocation set is partitioned into multiple mini-batches; gradients are computed sequentially for each mini-batch and accumulated in the parameter buffers, and a single optimizer step is performed only after all mini-batches have been processed. With the loss averaged consistently over the full batch (or, equivalently, by scaling each mini-batch loss by its relative batch size), each parameter update is numerically equivalent to a full-batch gradient step. This design eliminates performance discrepancies that would otherwise arise purely from different mini-batching strategies, ensuring that the reported differences reflect the modelling choices rather than optimization artefacts.

## S9 Effect of $n_b$ for accuracy of HIDON and DIDON

For both HIDON and DIDON, the error decreases rapidly when $n_b$ is small and then gradually saturates as $n_b$ increases. This is a typical signature of finite-dimensional solution-manifold approximation.

During PIDON training, the learned spatial bases $\{\varphi_i\}_{i=1}^{n_b}$ span a reduced subspace $\mathcal{V}_n$. HIDON and DIDON approximate the parametric solution manifold on the test set within $\mathcal{V}_n$; thus, generalisation is primarily determined by how well $\mathcal{V}_n$ represents the dominant solution variations across the parameter domain. When $n_b$ is too small, $\mathcal{V}_n$ is insufficient to capture these variations, leading to larger errors. As $n_\mathrm{b}$ grows, more independent spatial features are included and the error drops quickly. Beyond a certain $n_\mathrm{b}$, newly added bases contribute little additional coverage, so the error becomes weakly sensitive to $n_\mathrm{b}$ and saturates.

**S10 Definition of residual energy (RE) for quantifying subspace expressiveness**

To quantify the expressiveness of a learned subspace (i.e., its coverage of the solution manifold), we evaluate the residual energy (RE) of each solution snapshot after projection onto a prescribed basis subspace.

During PIDON training or POD, the learned spatial bases $\mathbf{\Phi} = \{\varphi_i\}_{i=1}^{n_\mathrm{b}}$ span a reduced subspace $\mathcal{V}_n$. We define the (Euclidean) orthogonal projector onto $\mathcal{V}_n$ as $P_\mathcal{V} = \mathbf{\Phi}(\mathbf{\Phi}^T\mathbf{\Phi})^{-1}\mathbf{\Phi}^T$. The residual energy (RE) of snapshot $u(\cdot;\mu)$ with respect to $\mathcal{V}_n$ is defined by

$$\mathrm{RE}(\mu,\mathcal{V}_n) := \frac{\left\|u(\cdot;\mu) - P_v u(\cdot;\mu)\right\|_2^2}{\left\|u(\cdot;\mu)\right\|_2^2}.$$

This quantity is a normalized squared projection error, hence $0 \le \mathrm{RE}(\mu,\mathcal{V}_n) \le 1$. A smaller RE indicates that the subspace $\mathcal{V}_n$ captures a larger fraction of the snapshot energy, i.e., provides stronger representational coverage of the solution manifold.

## References

[1] P. Benner, S. Gugercin, K. Willcox, A Survey of Projection-Based Model Reduction Methods for Parametric Dynamical Systems, SIAM Review, 57 (2015) 483-531.

[2] M. Yin, N. Charon, R. Brody, L. Lu, N. Trayanova, M. Maggioni, A scalable framework for learning the geometry-dependent solution operators of partial differential equations, Nature computational science, 4 (2024) 928-940.

[3] D. Huynh, Data‐driven physics‐based digital twins via a library of component‐based reduced‐order models, International Journal for Numerical Methods in Engineering, 123 (2022) 2986-3003.

[4] J.L. Lumley, The structure of inhomogeneous turbulent flows, Atmospheric turbulence and radio wave propagation, (1967) 166-178.

[5] L. Sirovich, Turbulence and the dynamics of coherent structures. I. Coherent structures, Quarterly of applied mathematics, 45 (1987) 561-571.

[6] G. Berkooz, P. Holmes, J.L. Lumley, The proper orthogonal decomposition in the analysis of turbulent flows, Annual review of fluid mechanics, 25 (1993) 539-575.

[7] C.W. Rowley, T. Colonius, R.M. Murray, Model reduction for compressible flows using POD and Galerkin projection, Physica D: Nonlinear Phenomena, 189 (2004) 115-129.

[8] G. Rozza, D.B.P. Huynh, A.T. Patera, Reduced basis approximation and a posteriori error estimation for affinely parametrized elliptic coercive partial differential equations: application to transport and continuum mechanics, Archives of Computational Methods in Engineering, 15 (2008) 229-275.

[9] M. Barrault, Y. Maday, N.C. Nguyen, A.T. Patera, An 'empirical interpolation'method: application to efficient reduced-basis discretization of partial differential equations, Comptes Rendus Mathematique, 339 (2004) 667-672.

[10] S. Chaturantabut, D.C. Sorensen, Nonlinear model reduction via discrete empirical interpolation, SIAM Journal on Scientific Computing, 32 (2010) 2737-2764.

[11] K. Carlberg, C. Farhat, J. Cortial, D. Amsallem, The GNAT method for nonlinear model reduction: effective implementation and application to computational fluid dynamics and turbulent flows, Journal of Computational Physics, 242 (2013) 623-647.

[12] B. Kramer, B. Peherstorfer, K.E. Willcox, Learning nonlinear reduced models from data with operator inference, Annual Review of Fluid Mechanics, 56 (2024) 521-548.

[13] Y. Kim, Y. Choi, D. Widemann, T. Zohdi, A fast and accurate physics-informed neural network reduced order model with shallow masked autoencoder, Journal of Computational Physics, 451 (2022) 110841.

[14] J.S. Hesthaven, S. Ubbiali, Non-intrusive reduced order modeling of nonlinear problems using neural networks, Journal of Computational Physics, 363 (2018) 55-78.

[15] K. Lee, K.T. Carlberg, Model reduction of dynamical systems on nonlinear manifolds using deep convolutional autoencoders, Journal of Computational Physics, 404 (2020) 108973.

[16] T. Chen, H. Chen, Universal approximation to nonlinear operators by neural networks with arbitrary activation functions and its application to dynamical systems, IEEE transactions on neural networks, 6 (1995) 911-917.

[17] L. Lu, P. Jin, G. Pang, Z. Zhang, G.E. Karniadakis, Learning nonlinear operators via DeepONet based on the universal approximation theorem of operators, Nature machine intelligence, 3 (2021) 218-229.

[18] S. Wang, H. Wang, P. Perdikaris, Learning the solution operator of parametric partial differential equations with physics-informed DeepONets, Science advances, 7 (2021) eabi8605.

[19] Z. Li, N. Kovachki, K. Azizzadenesheli, B. Liu, K. Bhattacharya, A. Stuart, A. Anandkumar, Fourier neural operator for parametric partial differential equations, arXiv preprint arXiv:2010.08895, (2020).

[20] G.E. Karniadakis, I.G. Kevrekidis, L. Lu, P. Perdikaris, S. Wang, L. Yang, Physics-informed machine learning, Nature Reviews Physics, 3 (2021) 422-440.

[21] W. Chen, Q. Wang, J.S. Hesthaven, C. Zhang, Physics-informed machine learning for reduced-order modeling of nonlinear problems, Journal of computational physics, 446 (2021) 110666.

[22] Y. Chen, S. Koohy, Gpt-pinn: Generative pre-trained physics-informed neural networks toward non-intrusive meta-learning of parametric pdes, Finite Elements in Analysis and Design, 228 (2024) 104047.

[23] Z. Li, H. Zheng, N. Kovachki, D. Jin, H. Chen, B. Liu, K. Azizzadenesheli, A. Anandkumar, Physics-informed neural operator for learning partial differential equations, ACM/IMS Journal of Data Science, 1 (2024) 1-27.

[24] S. Brivio, S. Fresca, A. Manzoni, PTPI-DL-ROMs: Pre-trained physics-informed deep learning-based reduced order models for nonlinear parametrized PDEs, Computer Methods in Applied Mechanics and Engineering, 432 (2024) 117404.

[25] S. Karumuri, L. Graham-Brady, S. Goswami, Physics-informed latent neural operator for real-time predictions of time-dependent parametric PDEs, Computer Methods in Applied Mechanics and Engineering, 450 (2026) 118599.

[26] J. Cho, S. Nam, H. Yang, S.-B. Yun, Y. Hong, E. Park, Separable physics-informed neural networks, Advances in Neural Information Processing Systems, 36 (2023) 23761-23788.

[27] L. Mandl, S. Goswami, L. Lambers, T. Ricken, Separable physics-informed DeepONet: Breaking the curse of dimensionality in physics-informed machine learning, Computer Methods in Applied Mechanics and Engineering, 434 (2025) 117586.

[28] B. Jacob, A.A. Howard, P. Stinis, SPIKANs: separable physics-informed Kolmogorov–Arnold networks, Machine Learning: Science and Technology, 6 (2025) 035060.

[29] Z. Liu, Y. Liu, X. Yan, W. Liu, H. Nie, S. Guo, C.-a. Zhang, Automatic network structure discovery of physics informed neural networks via knowledge distillation, Nature Communications, 16 (2025) 9558.

[30] A. Jiao, H. He, R. Ranade, J. Pathak, L. Lu, One-shot learning for solution operators of partial differential equations, Nature Communications, 16 (2025) 8386.

[31] S. Ingimarson, L.G. Rebholz, T. Iliescu, Full and reduced order model consistency of the nonlinearity discretization in incompressible flows, Computer Methods in Applied Mechanics and Engineering, 401 (2022) 115620.

[32] M.H. Stone, The generalized Weierstrass approximation theorem, Mathematics Magazine, 21 (1948) 237-254.

[33] A. Pinkus, Weierstrass and approximation theory, Journal of Approximation Theory, 107 (2000) 1-66.

[34] G. Cybenko, Approximation by superpositions of a sigmoidal function, Mathematics of control, signals and systems, 2 (1989) 303-314.

[35] K. Hornik, M. Stinchcombe, H. White, Multilayer feedforward networks are universal approximators, Neural networks, 2 (1989) 359-366.

[36] A. Pinkus, Approximation theory of the MLP model in neural networks, Acta numerica, 8 (1999) 143-195.

[37] S. Wang, Y. Teng, P. Perdikaris, Understanding and Mitigating Gradient Flow Pathologies in Physics-Informed Neural Networks, SIAM Journal on Scientific Computing, 43 (2021) A3055-A3081.

[38] A. Pinkus, N-widths in Approximation Theory, Springer Science & Business Media, 2012.

[39] T. Lassila, A. Manzoni, A. Quarteroni, G. Rozza, Generalized reduced basis methods and n-width estimates for the approximation of the solution manifold of parametric PDEs, in: Analysis and numerics of partial differential equations, Springer, 2013, pp. 307-329.

[40] J.S. Hesthaven, G. Rozza, B. Stamm, Certified reduced basis methods for parametrized partial differential equations, Springer, 2016.

[41] A. Ferrari, K. Willcox, Digital twins in mechanical and aerospace engineering, Nature Computational Science, 4 (2024) 178-183.

[42] R. Everson, L. Sirovich, Karhunen–Loeve procedure for gappy data, Journal of the Optical Society of America A, 12 (1995) 1657-1664.

[43] H. Wang, G. Jiang, W. Wang, Y. Liu, A novel hyper-reduction framework featuring direct projection without an approximation process, Physics of Fluids, 36 (2024).